\documentclass[twoside, a4paper, notoc]{JHEP3}
\usepackage{amsfonts}
\usepackage{amssymb}
\usepackage{epsfig}
\usepackage{graphicx}

\newcommand{\be}{\begin{equation}}
\newcommand{\ee}{\end{equation}}
\newcommand{\bea}{\begin{eqnarray}}
\newcommand{\eea}{\end{eqnarray}}
\newcommand{\mbb}{\mathbb}
\newcommand{\ti}{\times}
\newcommand{\half}{\frac{1}{2}}
\newcommand{\mc}{\mathcal}

\title{Astrophysical and Cosmological Implications of
Large Volume String Compactifications}
\author{Joseph P. Conlon, Fernando Quevedo\\ DAMTP, Centre for Mathematical Sciences,
  Wilberforce Road, Cambridge, CB3 0WA, UK.\\ E-mail: \email{j.p.conlon@damtp.cam.ac.uk, f.quevedo@damtp.cam.ac.uk}}

\abstract{ We study  the spectrum, couplings and
cosmological and astrophysical implications of the moduli fields for the class of
Calabi-Yau IIB string compactifications for which
moduli stabilisation leads to an
exponentially large volume $\mc{V} \sim 10^{15} l_s^6$ and an
intermediate string scale $m_s \sim 10^{11}\, \hbox{GeV}$, with
TeV-scale observable supersymmetry breaking.
 All K\"ahler moduli except for the overall
volume  are heavier than the susy breaking scale,
with $m \sim \ln(M_P/m_{3/2}) m_{3/2} \sim (\ln(M_P/m_{3/2}))^2
m_{susy}\sim 500 \, \hbox{TeV} $
and, contrary to
standard expectations, have matter couplings suppressed only by the
string scale rather than the Planck scale.
These decay to matter early in
the history of the universe, with a reheat temperature $T \sim 10^7 \hbox{GeV}$, and
 are free
from the cosmological moduli problem (CMP). The heavy moduli have a branching ratio to gravitino pairs of
$10^{-30}$ and do not suffer from the gravitino overproduction problem.
The overall volume  modulus is a
distinctive feature of these  models and is
an $M_{planck}$-coupled  scalar of mass $m \sim 1\, \hbox{MeV}$
and subject to the CMP. A period of thermal inflation may help relax
this problem. This field has a lifetime $\tau \sim 10^{24}$s and
can contribute to dark matter. It may be
detected through its decays to $\gamma \gamma$ or $e^+ e^-$.
If accessible the $e^+ e^-$ decay mode dominates, with $\rm{Br}(\chi \to \gamma \gamma)$ suppressed by a factor
$\left(\ln(M_P/m_{3/2})\right)^2$.
We consider the potential for  detection of this field
through different astrophysical sources: the Milky Way halo, the diffuse cosmic
background and nearby galaxy clusters and find that
the observed gamma-ray background constrains $\Omega_{\chi} \lesssim 10^{-4}$.
The decays of this field may generate the $511 \, \hbox{keV}$ emission line from the galactic centre observed by
INTEGRAL/SPI.}


\preprint{DAMTP-2007-43}

\begin{document}

\tableofcontents

\section{Introduction}

It is an old and hard problem to connect string compactifications to observational physics.
The principal difficulty is that the compactification energy scales
are usually much larger than those
directly accessible to experiment. However, we are helped by the fact that compactifications
do have generic and model-independent features.
One such feature is the presence of a moduli sector, consisting of many gravitationally coupled scalar fields.
String moduli are naively massless particles and as such would give rise to unobserved fifth forces. It is
therefore necessary that they receive a mass and the generation of flux-induced moduli potentials has been an active topic of research
over the last few years (for review articles see \cite{hepth0509003, hepth0610102, hepth0610327, hepth0701050}).

As the moduli determine the vacuum structure,
models with stabilised moduli are a prerequisite for doing string phenomenology.
One direction of research, looking towards particle physics,
 has been to study the structure of supersymmetry-breaking terms that arises, as such terms
can only be calculated once the vacuum has been identified. However, moduli can also play an
important role in cosmology. Open and closed string moduli
have recently been used to build inflation models within string theory.
Moduli tend to be good candidates for inflatons, as they are
flat prior to supersymmetry breaking and are ubiquitious in string
models as scalar fields which interact gravitationally and are
singlets under the standard model gauge group.
If sufficiently long-lived, moduli could also contribute to dark matter.
However, moduli also cause cosmological problems. Their
relatively weak, gravitational-strength interactions imply that moduli
 are either stable or decay late in the history of universe, and in the presence of low-energy
 supersymmetry
generic moduli either spoil nucleosynthesis or overclose
the universe.

It is helpful to re-examine late-time (i.e. post-inflationary) modular cosmology in the context of the explicit
models of moduli stabilisation that have been developed. Examples of work in this direction are
\cite{hepph0504036, hepph0602061, hepph0602081, hepph0604140, hepph0605091}.
In making contact with phenomenology one promising class of compactifications are the
large-volume models developed in \cite{hepth0502058, hepth0505076}. These
occur in flux compactifications of IIB string theory with D-branes and orientifold planes,
with the consistent inclusion of both $\alpha'$ and nonperturbative corrections.
These models dynamically stabilise the volume at
exponentially large values, allowing the generation of hierarchies.
The gravitino and string scales are given by
\be
m_{3/2} \sim \frac{M_P}{\mc{V}}, \qquad m_s \sim \frac{M_P}{\sqrt{\mc{V}}}.
\ee
Here $\mc{V}$ is the dimensionless volume - the physical volume is $\mc{V} l_s^6 \equiv \mc{V} (2 \pi \sqrt{\alpha'})^6$.
Thus a compactification volume of $10^{15} l_s^6$, corresponding to a string scale $m_s \sim 10^{11} \hbox{GeV}$, can generate
the weak hierarchy through TeV-scale supersymmetry \cite{biq}.
In these models other hierarchical scales also appear as different powers of the
volume - for example the axionic scale appears as $f_a \sim M_P/\sqrt{\mc{V}} \sim 10^{11} \hbox{GeV}$ \cite{hepth0602233}
and the neutrino suppression scale as $\Lambda \sim M_P/\mc{V}^{1/3} \sim 10^{14} \hbox{GeV}$ \cite{hepph0611144}.
We will give a more detailed review of large-volume models in section \ref{secLVM}.

The moduli for these models divide into two classes, $\Phi$ and $\chi$,
associated respectively with `small' cycles and the overall volume. These have masses
\be
m_{\Phi} \sim \ln(M_P/m_{3/2}) m_{3/2}, \qquad m_{\chi} \sim m_{3/2} \left(\frac{m_{3/2}}{M_P}\right)^{\half}.
\ee
The requirement of TeV supersymmetry constrains the mass of the light modulus to be $\sim 1 \hbox{MeV}$.
The purpose of this paper is to perform a detailed study of the physics and couplings of these
moduli, computing the decay modes and branching ratios.
We will see that starting with a well-motivated stringy construction, with a moduli potential that
naturally generates the weak hierarchy, gives results significant different from those obtained
under assumptions of generic behaviour \cite{hepph0602061, hepph0602081, hepph9701346, hepph9708226, hepph9802271}.
As a concrete example, the branching
ratio $\Phi \to \psi_{3/2} \psi_{3/2}$
is a factor $10^{30}$ smaller than the
$\mc{O}(1)$ expectations of \cite{hepph0602061, hepph0602081}.

The structure of this paper is as follows. In sections
\ref{secLVM} and \ref{sec3} we review the large-volume models and provide a precise computation of the masses
and couplings of the moduli fields.
These sections are more formal in nature and a reader more interested in the resulting
phenomenology of the moduli can skip these sections and start at section \ref{secModCosmo}, using the
results of section \ref{sec3} that are summarised in table \ref{table1}. In section \ref{secModCosmo} we review the cosmological
problems moduli can cause, while in section \ref{secLVMearly} we analyse the
behaviour of the large-volume moduli in the early universe and how they affect reheating, the cosmological moduli problem and
the gravitino overproduction problem. In section \ref{secLateUniverse} we study the ability of the moduli to contribute
to dark matter and examine the ability of the light modulus to contribute to the 511keV line.

This paper differs from
most of the recent literature on moduli cosmology, which has concentrated
on their potential role as inflatons. Here we will simply assume
that inflation has occurred in the early universe and concentrate on
the moduli cosmology in the post-inflationary era.

\section{Large Volume Models}
\label{secLVM}

Large volume models originate in string theory, but here
we view them simply as supergravity models. Their simplest avatar is that of
compactifications on $\mbb{P}^4_{[1,1,1,6,9]}$, which has two K\"ahler moduli, denoted by
$T_s = \tau_s + i b_s$ and $T_b = \tau_b + i b_b$. The `$s$' and `$b$' stand for `small' and `big'.
The Calabi-Yau volume is $\mc{V} = \frac{1}{9\sqrt{2}} \left( \tau_b^{3/2} - \tau_s^{3/2} \right)$ \cite{hepth0404257}.
The geometry should be thought of as analogous to a Swiss cheese - the small modulus controls the size of the
hole and the big modulus the size of the cheese.
In terms of these the K\"ahler potential and superpotential
are\footnote{In these string models there are also complex structure
  moduli $U$ and the dilaton $S$. Their scalar potential has been
  found to dominate at large volume unless they sit at their
  minimum \cite{hepth0502058}. This serves as a trapping mechanism for these fields. Even
  though they have masses of order the TeV scale and couple with
  gravitational strength \cite{hepth0505076}, this trapping indicates that while the
  K\"ahler moduli  roll through the scalar potential and could
  have coherent oscillations around their minima, the fields $U$ and
  $S$ energetically prefer to essentially sit at their minima and therefore do not cause a cosmological problem. In
  this note we will only study  the cosmological implications
  of the K\"ahler moduli.}
\bea
\label{KahlerPotential}
\mc{K} & = & - 2 \ln \left(\frac{1}{9\sqrt{2}}\left(\tau_b^{3/2} - \tau_s^{3/2}\right) + \frac{\xi}{2 g_s^{3/2}} \right) \\
\label{eqW}
W & = & W_0 + A_s e^{-a_s T_s}.
\eea
Here $\xi = \zeta(3) \chi(M)/(2 \pi)^3$ is a constant entering the $\alpha'$ correction (with $\chi(M)$
the Euler number of the Calabi-Yau manifold) and $g_s$ is the
string coupling. $W_0$ is $\mc{O}(1)$ and is the tree-level flux superpotential that arises after stabilising the dilaton and complex structure moduli.
For practical convenience in our computations, we will rewrite (\ref{KahlerPotential}) and (\ref{eqW}) as
\bea
\label{KahlerPotential2}
\mc{K} & = & - 2 \ln \left( \left( \tau_b^{3/2} - \tau_s^{3/2} \right) + \xi' \right) \\
\label{eqW2}
W & = & W_0 + A_s e^{-a_s T_s},
\eea
absorbing the overall factor of $9 \sqrt{2}$ into the value of $W_0$ and $A_s$ (so $W_0 \to 9\sqrt{2}W_0$ and
$A_s \to 9\sqrt{2} A_s$). Clearly this does not alter the physics in any way.
After extremising the axionic field, the supergravity scalar potential at large volumes is given by
\be
\label{potencial2}
V = \frac{ 8(a_s A_s)^2 \sqrt{\tau_s} e^{-2 a_s \tau_s}}{3 \tau_b^{3/2}} - \frac{4 a_s A_s W_0 \tau_s e^{-a_s \tau_s}}{\tau_b^3}
+ \frac{\nu \vert W_0 \vert^2}{\tau_b^{9/2}},
\ee
where $\nu = \frac{27 \sqrt{2} \xi}{4 g_s^{3/2}}$.

This potential has been studied in detail in \cite{hepth0502058, hepth0505076, 07040737}.
It has a non-supersymmetric AdS minimum at $\mc{V} \sim e^{a_s \tau_s} \gg 1$ with $\tau_s \sim \frac{\xi^{2/3}}{g_s}$.
This minimum has a negative cosmological constant of order $\frac{1}{\mc{V}^3}$. There exist various methods to introduce a
positive energy to uplift this minimum to de Sitter \cite{hepth0301240, hepth0309187, hepth0402135}, and the uplifted minimum
is stable against tunnelling \cite{07051557}. The physics presented in this paper
is not significantly affected by the details of the uplift, and so we do not consider the uplift further.

The stabilised exponentially large volume can generate hierarchies. As the gravitino mass is given by
\be
\label{qqq}
m_{3/2} = e^{\hat{K}/2}W = \frac{W_0}{\mc{V}},
\ee
it follows that an exponentially large volume can lead to a gravitino mass exponentially lower than the Planck scale.
This allows a natural solution of the hierarchy problem through TeV-scale supersymmetry breaking.
It follows from (\ref{qqq}) that a TeV-scale gravitino mass requires $\mc{V} \sim 10^{15}$.
Through a detailed analysis
of the moduli potential and the F-terms that are generated \cite{hepth0610129},
it can in fact be shown that the scale of soft terms is lowered compared to the gravitino mass by a factor
$\ln(M_P/m_{3/2})$, so
$$
m_{soft} = \frac{m_{3/2}}{\ln(M_P/m_{3/2})}.
$$
A sensible phenomenology therefore requires $m_{3/2} \sim 20 \hbox{TeV}$.

The potential (\ref{potencial2}) generates masses for the moduli.
Estimates of these masses
can be computed using $m_b^2 \sim \mc{K}_{bb}^{-1} \partial^2 V/\partial \tau_b^2$ and
$m_s^2 \sim \mc{K}_{ss}^{-1} \partial^2 V / \partial \tau_s^2$, giving\footnote{The axionic partners of $\tau_b,
  \tau_s$ also receive masses after their stabilisation. The partner
  of $\tau_s$ has a mass of the same order as $\tau_s$ whereas the
  axionic partner of $\tau_b$ is essentially massless. Being an axion, it does not
  couple directly to observable matter and therefore does not play a
  role in our cosmological discussion below.}
$$
m_{\tau_b} \sim \frac{M_P}{\mc{V}^{3/2}}, \qquad m_{\tau_s} \sim \frac{M_P \ln (M_P/m_{3/2})}{\mc{V}}.
$$
The light field is associated with the modulus controlling the overall volume,
whereas the heavy field is that associated with the small blow-up cycle.
In section \ref{sec3} we give a much more detailed analysis of the spectrum of moduli masses and couplings.

\section{Moduli Properties and Couplings}
\label{sec3}

In this section we describe how to canonically normalise the moduli and compute their masses
and couplings to matter particles.

\subsection{Normalisation and Couplings to Photons}

We assume the minimum of the moduli potential has been located.
By writing $\tau_i = \langle \tau_i \rangle + \delta \tau_i$, we can always expand the
Lagrangian about the minimum of the moduli potential. In the vicinity
of the minimum, we can write
\be
\label{lagfull}
\mc{L} = \mc {K}_{i \bar{j}} \partial_\mu (\delta \tau_i) \partial^\mu (\delta \tau_j) - V_0 - (M^2)_{i j} (\delta \tau_i)(\delta \tau_j) -
\mc{O}(\delta \tau^3)
- \kappa \frac{\tau_{\alpha}}{M_P} F_{\mu \nu} F^{\mu \nu}.
\ee
Here we take $f_{U(1)} = \kappa \tau_{\alpha}$ where $\kappa$ is a
normalisation constant and  $\alpha$ labels one of the
small four-cycles since we assume the
standard model lives on a stack of D7 branes wrapping the small four-cycle.\footnote{A
D7 wrapping the large four-cycle would give rise to unrealistically
small values of the gauge couplings ($1/g^2\sim \mc{V}^{2/3} \sim 10^{10}$).}
 To express the Lagrangian
(\ref{lagfull}) in terms of canonically normalised fields,
we require the eigenvalues and normalised eigenvectors of $(\mc{K}^{-1})_{i\bar{j}} (M^2)_{\bar{j}k}$.
Anticipating our use of the $\mbb{P}^4_{[1,1,1,6,9]}$ model, we now specialise to a
2-modulus model, in which we denote  $\tau_1\equiv \tau_b,
\tau_2\equiv \tau_s$. This sets $\alpha=2=s$ above. In this case we write the eigenvalues and eigenvectors of
$(\mc{K}^{-1})_{i\bar{j}} (M^2)_{\bar{j}k}$ as
$m_{\Phi}^2, m_{\chi}^2$, and $v_{\Phi}, v_{\chi}$ respectively, with $m_{\Phi} > m_{\chi}$. The eigenvectors are normalised
as $v_{\alpha}^T \cdot \mc{K} \cdot v_{\beta} = \delta_{\alpha \beta}$.

We may rewrite the Lagrangian in
terms of canonical fields $\Phi$ and $\chi$ defined by
\be
\label{uno}
\left( \begin{array}{c} \delta \tau_b \\ \delta \tau_s \end{array} \right) = \Bigg( v_\Phi \Bigg) \frac{\Phi}{\sqrt{2}}
+ \Bigg( v_\chi \Bigg) \frac{\chi}{\sqrt{2}}.
\ee
Canonically normalising the $U(1)$ kinetic term, the Lagrangian (\ref{lagfull}) can be written as
$$
\mc{L} = \half \partial_\mu \Phi \partial^\mu \Phi + \half \partial_\mu \chi \partial^\mu \chi - V_0
- \half m_{\Phi}^2 \Phi^2 - \half m_{\chi}^2 \chi^2 - \frac{1}{4} F_{\mu \nu}
F^{\mu \nu} -
\frac{ \left( \Phi (v_\Phi)_s  + \chi (v_{\chi})_s \right) }{4 \sqrt{2} \langle \tau_s \rangle M_P}
F_{\mu \nu} F^{\mu \nu}.
$$
The coupling of the two moduli $\Phi$ and $\chi$ to photons, which we denote by $\lambda$, is then given by
\bea
\label{llambda}
\lambda_{\Phi \gamma \gamma} & = & \frac{ (v_\Phi)_s  }{\sqrt{2} \langle \tau_s \rangle}, \nonumber \\
\lambda_{\chi \gamma \gamma} & = &  \frac{ (v_{\chi})_s }{\sqrt{2} \langle \tau_s \rangle}.
\eea
Thus, given the moduli Lagrangian we can follow a well-defined procedure to
compute the moduli couplings to photons.

The explicit forms of the matrices $(\mc{K}^{-1})_{i\bar{j}}$ and
$(M^2)_{\bar{j}k}$ for the large volume models can be computed and are given in the appendix.
Importantly, it follows from the
expression for the moduli K\"ahler potential (\ref{KahlerPotential}) that there is a small mixing
between the moduli $\tau_b$ and $\tau_s$, and the canonically normalised fields couple to matter living
on both small and large cycles.
The matrix $\mc{K}^{-1} M^2$ takes the form:
\be
\label{matriz}
\mc{K}^{-1} M^2 = \frac{2 a_s\langle \tau_s \rangle |W_0|^2 \nu}{3 \langle \tau_b \rangle^{9/2} } \left( \begin{array}{ccc}
  -9(1-7\epsilon) & & 6 a_s \langle \tau_b \rangle(1-5\epsilon+16\epsilon^2)\\
-\frac{6\langle \tau_b \rangle^{1/2}}{\langle \tau_s \rangle^{1/2}} (1-5\epsilon+4\epsilon^2) & &  \, \,\,\,
\frac{4 a_s \langle \tau_b \rangle^{3/2}}{\langle \tau_s \rangle^{1/2}}  (1-3\epsilon+6\epsilon^2) \end{array} \right),
\ee
where $\epsilon = (4 a_s \langle \tau_s \rangle)^{-1}$ and the expressions are valid to
 $\mc{O}(\epsilon^2)$ (there are also $1/\mc{V}$
 corrections, which are negligible). (\ref{matriz}) has one large and one small eigenvalue, denoted by
 $m_{\Phi}^2$ and $m_{\chi}^2$. Because $m_{\Phi}^2 \gg m_{\chi}^2$, we have at leading order in $\epsilon$:
\bea
\label{massp}
m_\Phi^2\ &  \simeq  & \, \hbox{Tr} \left( \mc{K}^{-1}M^2\right)\,
\simeq \, \frac{8 \nu
  |W_0|^2 a_s^2 \langle \tau_s \rangle^{1/2}}{3\langle \tau_b \rangle^3}\, 
  = (2 m_{3/2} \ln(M_P/m_{3/2}))^2 \, \sim \left(\frac{\ln \mc{V}}{\mc{V}}\right)^2 \\
\label{massc}
m_\chi^2 \ & \simeq  & \, \frac{\hbox{Det}\left(
    \mc{K}^{-1}M^2\right)}{\hbox{Tr}\left(\mc{K}^{-1}M^2\right)}
 \,  \simeq \, \frac{27 |W_0|^2 \nu}{4 a_s \langle \tau_s \rangle \langle \tau_b \rangle^{9/2}}\,
 \,  \sim \,   \mc{V}^{-3}/\ln\mc{V}.
\eea
We can see  explicitly  the large hierarchy of masses among the two
 observable particles, with $\Phi$ heavier than the gravitino mass and
$\chi$ lighter by a factor of $\sqrt{\mc{V}}$. We have numerically confirmed the analytic mass formulae of (\ref{massp}) and
(\ref{massc}).\footnote{Numerically, the effect of including an
uplifting potential $\delta V \sim \frac{\epsilon}{\mc{V}^2}$
is to reduce $m_{\chi}$ from the value given in (\ref{massc}), $m_{\chi} \to 0.6 m_{\chi}$, while leaving
$m_{\Phi}$ unaffected.}

Finding the eigenvectors of $\mc{K}^{-1}M^2$ and using (\ref{uno}) we can write the original
fields $\delta\tau_{b,s}$ in terms of $\Phi$ and $\chi$ (in Planck
units) as:\footnote{Notice that since the
light field $\chi$ is dominantly the volume modulus, for which the
K\"ahler potential
can be approximated by
$K = - 3 \ln (T_b + \bar{T}_b)$. In this case one can perform the
canonical normalisation
for all values of the field, obtaining
$
\frac{\delta \tau_b}{\tau_b} = \sqrt{\frac{2}{3}} \chi.
$
This is precisely the coefficient we find in equation (\ref{eqs})}
\bea
\label{eqs}
\delta \tau_b & = &
\left(\sqrt{6}\langle \tau_b \rangle^{1/4}\langle \tau_s \rangle^{3/4}\left(1-2\epsilon\right)\right)\,
\frac{\Phi}{\sqrt{2}} + \left(\sqrt{\frac{4}{3}} \langle \tau_b \rangle\right)\,   \frac{\chi}{\sqrt{2}}\,  \sim\, \mc{O}\left( {\mc V}^{1/6}\right)\,
\Phi\, +\, \mc{O}\left({\mc V}^{2/3}\right)\, \chi\nonumber \\
\\
\delta \tau_s & = & \left(\frac{2\sqrt{6}}{3} \langle \tau_b \rangle^{3/4}\langle \tau_s \rangle^{1/4} \right)\,
 \frac{\Phi}{\sqrt{2}} +
 \left(\frac{\sqrt{3}}{a_s}\left(1-2\epsilon\right) \,\right) \frac{\chi}{\sqrt{2}} \, \sim \,
      \mc{O}\left(\mc{V}^{1/2}\right)\, \Phi\, +\,
      \mc{O}\left(1\right)\,  \chi \nonumber
\eea

This shows, as expected, that $\tau_b$ is mostly $\chi$ and $\tau_s$
is mostly $\Phi$. However there is an important mixing, which is subleading and has
coefficients depending on different powers of the volume $\mc{V}$.
This illustrates the fact that although the large modulus $\tau_b$ has no couplings to
photons, the light field $\chi$, although mostly aligned with $\tau_b$,
does have a measurable coupling to photons due to its small component
in the $\tau_s$ direction. This $\chi \gamma \gamma$ coupling is determined by the coefficient
$\frac{\sqrt{6}}{2a_s}$ in (\ref{eqs}), which happens to be volume independent.

The $\chi$ Lagrangian is therefore
\be
\label{photoncoupling}
\mc{L}_\chi = -\half \partial_\mu \chi \partial^\mu \chi - \half m_{\chi}^2 \chi^2 - \frac{1}{4} F_{\mu \nu} F^{\mu \nu}
- \frac{1}{4} \left( \frac{\sqrt{6}}{2 a_s \langle \tau_s \rangle} \right) \frac{\chi}{M_P} F_{\mu \nu} F^{\mu \nu}.
\ee
The Planck mass dependence is here included for explicitness.
Notice that the coupling of $\chi$ to photons is not only suppressed
by the Planck scale $M_P$, as one might naively expect, but it also
has a further suppression factor
proportional to
 \be
a_s \langle \tau_s \rangle \sim {\ln\left(M_p/m_{3/2}\right)}
 \sim {\ln\mc{V}}.
\ee
The dimensionful coupling of $\chi$ to photons is
\be
\label{cggcoup}
\lambda_{\chi\gamma\gamma} = \frac{\sqrt{6}}{2 M_P\ln\left(M_P/m_{3/2}\right)},
\ee
 and so it is slightly weaker than standard moduli
couplings to matter. Naively one might have supposed a purely
Planckian coupling, with $\lambda_{\chi\gamma\gamma} = 1/M_P$
 (as done in \cite{hepph9701346, hepph9708226, hepph9802271}).
 We see that the result in a more realistic model actually suppresses the decay rate by a factor
 of $\ln(M_P/m_{3/2})^2\sim 1000$. This suppression of the $2 \gamma$ decay mode
 will subsequently play an important role when we discuss the possible role of $\chi$ in generating the
 511keV line from the galactic centre.

From (\ref{eqs}) it also follows that  the photon couplings to the heavy field $\Phi$ will
involve a factor $\mc{V}^{1/2}$ rather than
$\frac{\sqrt{6}}{2a_s}$. The dimensionful coupling is
\be
\lambda_{\Phi\gamma\gamma} \sim
\left(\frac{2}{\sqrt{3}} \frac{ \langle \tau_b \rangle^{3/4}}{\langle \tau_s \rangle^{3/4} M_P} \right)\,
\sim \frac{\sqrt{\mc{V}}}{M_P} \sim \, \frac{1}{m_s}.
\ee
This implies that the interactions of $\Phi$ with photons are only
suppressed by the string scale $m_s \ll M_P$ rather than the Planck
scale and therefore the decay rates of the heavy fields $\Phi$ are
much faster than is usually assumed for moduli fields. As we will explore later, this
feature is crucial when studying the behaviour of these fields in the early universe.

\subsection{Couplings to Electrons}

Here we compute the magnitude of the modular couplings to
$e^{+}e^{-}$.  This arises from the supergravity Lagrangian, with
the relevant terms being
\bea
\mc{L} & = & K_{\bar{e} e} \bar{e} \gamma^\mu \partial_\mu e + K_{H \bar{H}} \partial_\mu H \partial^\mu \bar{H} +
e^{\mc{K}/2} \partial_i \partial_j W \psi^i \psi^j, \nonumber \\
& = & K_{\bar{e} e} \bar{e} \gamma^\mu \partial_\mu e + K_{H \bar{H}} \partial_\mu H \partial^\mu \bar{H} +
e^{\mc{K}/2} \lambda H \bar{e} e.
\eea
To proceed we need to
know the K\"ahler metric for the chiral matter fields.
We use the result \cite{hepth0609180}
\be
K_{\bar{e} e} \sim K_{\bar{H} H} \sim \frac{\tau_s^{1/3}}{\tau_b} = K_0
\left(1 + \frac{1}{3} \frac{\delta \tau_s}{\langle\tau_s\rangle}
 - \frac{\delta \tau_b}{\langle\tau_b\rangle} + \ldots \right) .
\ee
where $K_0\equiv \left\langle \frac{\tau_s^{1/3}}{\tau_b}
\right\rangle =\frac{\langle \tau_s \rangle^{1/3}}{\langle \tau_b \rangle}$.
We also need the expansion
\be
e^{\mc{K}/2} = \frac{1}{\mc{V}} \sim \frac{9\sqrt{2}}{\tau_b^{3/2} - \tau_s^{3/2}}
= \frac{1}{\mc{V}_0} \left( 1 - \frac{3}{2} \left( \frac{\delta \tau_b}{\langle \tau_b \rangle} \right) + \ldots \right),
\ee
where $\mc{V}_0=\langle\mc{V}\rangle$. The Lagrangian is then
\bea
\mc{L} & = & K_{0}\, \bar{e} \gamma^\mu \partial_\mu e + K_{0}\, \partial_\mu H \partial^\mu \bar{H} +
\frac{1}{\mc{V}_0}\lambda H \bar{e}e + \left( \frac{1}{3} \left(\frac{\delta \tau_s}{\langle \tau_s \rangle} \right)
- \left( \frac{\delta \tau_b}{\langle \tau_b \rangle} \right) \right) K_{0}\, \bar{e} \gamma^\mu \partial_\mu e \nonumber \\
& & + \left( \frac{1}{3} \left(\frac{\delta \tau_s}{\langle \tau_s \rangle} \right)
- \left( \frac{\delta \tau_b}{\langle \tau_b \rangle} \right) \right) K_{0}\, \partial_\mu H \partial^\mu \bar{H}
- \frac{3}{2} \left( \frac{\delta \tau_b}{\langle \tau_b \rangle} \right) \frac{1}{\mc{V}_0} \lambda H \bar{e} e.
\eea

We can now canonically normalise the matter fields and impose electroweak symmetry breaking, giving the Higgs a vev
and generating the electron mass. The effective Lagrangian for the electron field is
\be
\label{elecLag}
\bar{e}\, (\gamma^\mu \partial_\mu + m_e)\,  e + \left( \frac{1}{3} \frac{\delta \tau_s}{\langle \tau_s \rangle}
-  \frac{\delta \tau_b}{\langle \tau_b \rangle}  \right) \bar{e}\,
(\gamma^\mu \partial_\mu + m_e)\,  e
- \left( \frac{1}{3}  \frac{\delta \tau_s}{\langle \tau_s \rangle}  + \half  \frac{\delta \tau_b}{\langle \tau_b \rangle}  \right)
m_e\,  \bar{e} e.
\ee
The second term of (\ref{elecLag}) does not contribute to the $\chi$ decay rate - for onshell final-state particles the Feynman
amplitude vanishes due to the equations of motion.
The physical decay rate is determined by the final term of (\ref{elecLag}),
\be
\label{dre}
\frac{1}{3} \left( \frac{\delta \tau_s}{\langle \tau_s \rangle} \right) + \half \left( \frac{\delta \tau_b}{\langle \tau_b \rangle} \right),
\ee
and in particular how this converts into a linear combination of
$\Phi$ and $\chi$. 
Using the expression (\ref{eqs}) we obtain
\be
\label{eecoupling}
\delta\mc{L}_{\chi ee}\, \sim\,  \left( 1 + \frac{{1}}{a \langle \tau_s \rangle} \right) \, \frac{1}{\sqrt{6}} \frac{\chi}{M_P} m_e \bar{e}  e.
\ee
This is dominated by the former term, arising from the alignment of $\chi$ with the overall volume direction.
The coupling (\ref{eecoupling}) is suppressed by the Planck scale, but unlike (\ref{photoncoupling}) there
is no further parametric suppression.

For the heavy field $\Phi$, we find, similar to the couplings to photons, that the important term in (\ref{dre}) is the
$\frac{\delta \tau_s}{\langle \tau_s \rangle}$ term. Using the expansion (\ref{eqs}) we again see that the coupling of
$\Phi$ to electrons is suppressed only by the string scale rather than by the Planck scale:
\be
\delta\mc{L}_{\Phi ee}\, \sim\, \frac{\sqrt{\mc{V}}\chi}{M_P} m_e \bar{e} e \sim \frac{\chi}{m_s} m_e \bar{e} e.
\ee

\subsection{Computation of Moduli Lifetimes}
\label{lifetimes}

We now use the results of the previous sections to compute the moduli lifetimes.
After canonical normalisation we always obtain a Lagrangian
\be
\label{lag}
\mc{L} = - \frac{1}{4} F_{\mu \nu} F^{\mu \nu} - \frac{1}{2} \partial_\mu \phi
\partial^\mu \phi - \half m_\phi^2 \phi^2 + \frac{\lambda \phi}{4 M_P} F_{\mu \nu} F^{\mu \nu}
+ \mu \frac{\phi}{M_P} \bar{e} e.
\ee
Here $\phi$ represents either of the fields $\Phi, \chi$.
In terms of $m_\phi$, $\lambda$ and $\mu$, it is straightforward to compute
the $\phi$ decay rates, which are given by
\bea
\label{gammadec}
\Gamma_{\phi \to \gamma \gamma} & = & \frac{\lambda^2 m_\phi^3}{64 \pi M_P^2}, \\
\label{eedec}
\Gamma_{\phi \to e^{+} e^{-} } & = &  \frac{\mu^2 m_e^2 m_{\phi} }{8 \pi M_P^2} \left( 1 - \frac{4 m_e^2}{m_{\phi}^2} \right)^{3/2}.
\eea
The lifetimes for each decay mode are $\tau = \Gamma^{-1}$.
Using $M_P^{-1} = (2.4 \ti 10^{18} \hbox{GeV})^{-1} = 2.7 \ti 10^{-43}\rm{s},$
we can write:
\bea
\label{lif}
\tau_{\phi \to \gamma \gamma} & = & \frac{7.5 \ti 10^{23}\rm{s}}{\lambda^2} \left( \frac{1 \hbox{MeV}}{m_\phi} \right)^3, \\
\label{lifee}
\tau_{\phi \to e^{+} e^{-}} & = & \frac{3.75 \ti 10^{23}\rm{s}}{\mu^2} \left( \frac{1 \hbox{MeV}}{m_{\phi}} \right)
\left( 1 - \left( \frac{1 \hbox{MeV}}{m_{\phi}} \right)^2 \right)^{-3/2}.
\eea
For the light modulus $\chi$, substituting  $\lambda$ by
$\lambda_{\chi\gamma\gamma}\sim 1/\ln(M_P/m_{3/2}) \sim 0.038$ given in equation
(\ref{llambda}) and $m_\chi\sim 2$ MeV,  we have
\bea
\tau_{\chi \to \gamma \gamma} & \sim & 6\times 10^{25}\, \rm{s}, \\
\tau_{\chi \to e^{+}e^{-}} & \sim & 1.7 \ti 10^{24} \, \rm{s},
\eea
which is much larger than the age of the universe $\sim 3\times
10^{17}\,  \rm{s}$. From (\ref{lifee}) we can see that for $m_{\chi} \gtrsim 1 \hbox{MeV}$ the decay
to $e^{+} e^{-}$ pairs is dominant, with a branching ratio $\sim 0.97$.

For the heavy modulus $\Phi$, we have $\lambda_{\Phi\gamma\gamma}\sim
\sqrt{\mc{V}} \sim 10^7$ and $m_\Phi \sim 1000$ TeV. We then obtain
\be
\label{philifetime}
\tau_{\Phi} \sim 10^{-17}\rm{s},
\ee
which means the heavy moduli decay very early in the history of the universe.
The moduli lifetimes differ by a factor $\sim 10^{43}$: this large discrepancy originates in the
very different masses and couplings of the two moduli.

\subsection{Couplings and Decays to Gravitini}

Another decay mode of interest is that to gravitini. This mode is interesting because of the
danger of overproducing gravitini from moduli decays that give rise to reheating. While this mode
is inaccesible for the light modulus $\chi$, for the heavy field $\Phi$ this mode is present.
In \cite{hepph0602061, hepph0602081} it was shown that for many models with heavy moduli, the gravitino branching ratio
for moduli is $\mc{O}(1)$. This causes severe cosmological problems, as the decays of
such gravitini either spoil nucleosynthesis or overproduce supersymmetric dark matter.
However, for large volume models the branching ratio is negligible:
the gravitino is a bulk mode, while the heavy
modulus is located on the small cycle.
While the couplings of the heavy modulus to matter are suppressed by the string scale, those
to the gravitino are suppressed by the Planck scale.

For example, we can consider the $\Phi \to 2 \psi_{3/2}$ decay channel analysed in
\cite{hepph0602061, hepph0602081}. This arises from the Lagrangian term
\bea
\mc{L} & \sim & e^{G/2} \bar{\psi}_\mu \left[ \gamma^\mu, \gamma^\nu \right] \psi_\nu \nonumber \\
& = & e^{G/2} \Big( \left( \partial_{\tau_s} G \right) (\delta \tau_s) + \left( \partial_{\tau_b} G \right)
(\delta \tau_b) \Big) \bar{\psi}_\mu \left[ \gamma^\mu, \gamma^\nu \right] \psi_\nu
\eea
Here $G = \mc{K} + \ln W + \ln \bar{W}$. We now relate $\delta \tau_s$ and $\delta \tau_b$ to $\Phi$ and $\chi$ using
(\ref{eqs}), and use the fact that $\partial_{\tau_s} G \sim \frac{1}{\mc{V}}$, $\partial_{\tau_b} G \sim \frac{1}{\mc{V}^{2/3}}$, to
get
\bea
\label{phigrav}
\mc{L} & \sim & m_{3/2} \left( \frac{1}{\mc{V}} \left( \sqrt{\mc{V}} \Phi + \chi \right) + \frac{1}{\mc{V}^{2/3}}
\left( \mc{V}^{1/6} \Phi + \mc{V}^{2/3} \chi \right) \right) \bar{\psi}_\mu \left[ \gamma^\mu, \gamma^\nu \right] \psi_\nu \nonumber \\
& \sim & \left( \frac{1}{\sqrt{\mc{V}}} \frac{\Phi}{M_P} + \frac{\chi}{M_P} \right) m_{3/2} \bar{\psi}_\mu \left[ \gamma^\mu, \gamma^\nu \right] \psi_\nu.
\eea
The Lagrangian term
$$
\mc{L} \sim \epsilon^{\mu \rho \sigma \tau} \sum \left( (\partial_{T_i} G) \partial_\rho T_i - (\partial_{\bar{T}_i} G) \partial_\rho
\bar{T}_i \right) \bar{\psi}_{\mu} \gamma_\nu \psi_{\sigma},
$$
here only generates an axion-gravitino coupling and does not contribute to the $\Phi$ decay rate.
From (\ref{phigrav}), we then find
\be
\Gamma_{\Phi \to 2 \psi_{3/2}}
 \sim
 \frac{1}{\mc{V}} \frac{m_{\Phi}^3}{M_P^2},
\ee
where we have focused on the dominant volume scaling. As
\be
\Gamma_{\Phi \to e^{+}e^{-}} \sim \mc{V} \frac{m_{\Phi}^3}{M_P^2},
\ee
(see (\ref{gammadec}) and (\ref{philifetime}) above),
this implies that the branching ratio for gravitino pair production is
$\rm{Br}(\Phi \to 2 \psi_{3/2}) \sim \mc{V}^{-2} \sim 10^{-30}$!

The striking contrast between this result and
the $\mc{O}(1)$ branching ratios found in \cite{hepph0602061, hepph0602081} is that for the large-volume models there exists a
 double suppression: first,
the gravitino is a bulk mode which gives a suppression $\left( \frac{m_s}{M_P} \right)^2 = \mc{V}^{-1}$, and secondly, the dominant F-term
(again by a factor of $\mc{V}$) is that associated with the light overall volume modulus rather than the small heavy
modulus.\footnote{We stress however that it is still the F-term $F^{\Phi}$ that determines the physical soft terms, due to the much stronger matter couplings
of $\Phi$ than $\chi$ ($m_s^{-1}$ rather than $M_P^{-1}$).}
The $\Phi \to 2 \psi_{3/2}$ decay mode is therefore suppressed by a factor $\sim \mc{V}^2 \sim 10^{30}$ compared to the results of
\cite{hepph0602061, hepph0602081}.

In table \ref{table1} we summarise the results of this section for the properties, couplings and
decay modes of the moduli. In the
next sections we will examine the cosmological and astrophysical applications of these results.

\TABLE{\label{table1}
\begin{tabular}{|c|c|c|}
\hline
 & Light modulus $\chi$ & Heavy Modulus $\Phi$ \\
\hline
Mass & $\sim m_{3/2} \left(\frac{m_{3/2}}{M_P}\right)^{\half} \sim 2 \hbox{MeV}$ & 2 $m_{3/2} \ln(M_p/M_{3/2}) \sim 1200 \hbox{TeV}$ \\
\hline
Matter Couplings & \qquad \qquad $M_P^{-1}$ \phantom{trons} (electrons) & $m_s^{-1}$ \\
& $ \left( M_P \ln \left( \frac{M_P}{m_{3/2}} \right) \right)^{-1}$ (photons) & \\
\hline
Decay Modes & & \\
$\gamma \gamma$  & $\rm{Br} \sim 0.025, \qquad  \tau \sim 6.5 \ti 10^{25}$s &$\phantom{,} \rm{Br} \sim \mc{O}(1), \qquad \phantom{,,,} \tau \sim 10^{-17}$s\\
$e^{+} e^{-}$ & $\rm{Br} \sim 0.975, \qquad \tau \sim 1.7 \ti 10^{24}$s & $\phantom{,} \rm{Br} \sim \mc{O}(1), \qquad \phantom{,,,} \tau \sim 10^{-17}$s\\
$q \bar{q}$ & \rm{inaccessible} & $\phantom{,} \rm{Br} \sim \mc{O}(1),\phantom{,,,} \qquad \tau \sim 10^{-17}$s \\
$\psi_{3/2} \psi_{3/2}$ & \rm{inaccessible} & $\rm{Br} \sim 10^{-30},\phantom{,,} \qquad \tau \sim 10^{13}$s \\
\hline
\end{tabular}
\caption{The properties of the two moduli and their decay modes. The lifetimes quoted are for
sample masses of $m_{\Phi} = 1200 \hbox{TeV}$ and $m_{\chi} = 2 \hbox{MeV}$, with a string scale of $m_s = 10^{11} \hbox{GeV}$ and
a gravitino mass of 20 TeV. The scale of soft terms here is $m_{3/2}/\ln(M_P/m_{3/2}) \sim 500 \hbox{GeV}$.}}

\section{Review of Moduli Cosmology}
\label{secModCosmo}

As mentioned in the introduction, moduli fields have been widely
studied as possible candidates for inflation. There are currently
several competing scenarios in which the inflaton is either
an open string modulus or a closed string modulus.
In particular, for the large volume models there exists a natural
mechanism to generate a flat potential for one of the
`small' K\"ahler moduli as long as the Calabi-Yau has more than three
K\"ahler moduli \cite{hepth0509012}. This scenario has been further studied in
\cite{hepth0603246, hepth0606089, hepth0612197, 07040212} where more
inflationary trajectories were identified. There is a potential danger
that extra quantum corrections to the K\"ahler potential could spoil the
slow roll. However, this inflationary scenario also requires a string
scale of order the GUT scale in order to achieve the correct COBE
normalisation for the density perturbations. This is in tension with the scales required for
particle physics, as the GUT string scale gives a very heavy gravitino
$\sim 10^{13} \hbox{GeV}$ (cf \cite{KalloshLinde, realistic}) 
which is incompatible with low-energy supersymmetry. This is an
interesting challenge that may need  realisations of inflation at a
low scale \cite{LowScale} or a dynamical change in the volume after
inflation
in order  to 
satisfy the low-energy phenomenological requirements.

For our purpose we will assume the string scale in our vacuum is the intermediate scale $10^{11} \hbox{GeV}$ as preferred
by particle physics.
We   leave as an open problem to
develop a successful scenario of inflation within the context of the
intermediate scale models that we consider here.
Here we will simply assume that such an inflationary scenario can be developed
and concentrate on the subsequent cosmological evolution after inflation,
with the K\"ahler moduli rolling along their potential.
Over the years moduli have been associated with
several cosmological problems. Let us summarise the main issues.

\subsection{Cosmological Moduli Problem}

It is well-known that generic moduli with a mass $m \lesssim 1 \hbox{TeV}$ pose
problems for early-universe cosmology \cite{Coughlan, hepph9308292, hepph9308325}.
Such moduli masses are unavoidable in the conventional picture of gravity-mediated supersymmetry breaking,
where moduli obtain masses
comparable to the supersymmetry breaking scale, $m_{\phi} \sim m_{3/2} \sim m_{susy}$.
In gauge-mediated models, the problem is even more serious as the moduli masses are then lower than the
supersymmetry breaking scale, $m_{\phi} \sim m_{3/2} \ll m_{susy}$.
The problem is that the moduli are long-lived and after inflation come to dominate the energy density of the
universe.

This is a serious and model independent problem for light
scalar fields that couple gravitationally. Let us briefly review the source of this
problem.
We assume a scalar field $\phi$ with gravitational strength
interactions in a FRW background. Its time evolution is governed by the
equation\footnote{Strictly this applies to the time-averaged amplitude of the field oscillations.}
\be
\label{tevol}
\ddot{\phi}\ + \left(3H + \Gamma_\phi \right) \dot{\phi} \ +
\frac{\partial{V}}{\partial\phi}\ =\ 0,
\ee
where $H=\frac{\dot{a}}{a}$ is the Hubble parameter,
$a$ the scale factor, $V$ the
scalar potential and $\Gamma_\phi \sim  m_\phi^3/M_P^2$
 the $\phi$ decay rate.
Due to its original
supersymmetric flat potential, it is expected that after
inflation the modulus is not at its zero-temperature minimum
but instead at some initial value $\phi_{in}\sim M_P$. While
$t<t_{in}\sim m_\phi^{-1}$,
$H>m_\phi$ and the friction term $3H\dot{\phi}$ dominates the time
evolution of $\phi$, causing $\phi$ to remain at $\phi\sim \phi_{in}$.
At $t>t_{in}$ when the universe
is at a temperature $T_{in} \sim \sqrt{m_\phi M_P}$
(since the Friedmann equation implies $H \sim T^2/M_P$ for radiation), the field starts
oscillating around its minimum. Coherent oscillations
of the field after this time will come to dominate the energy density of the
universe since the initial energy density $\rho_\phi(T_{in})\sim
m_\phi^2\phi_{in}^2$
 increases with respect to standard radiation
density. The reason is that energy in coherent oscillations decreases
with $a^{-3}$  \cite{preskill}
whereas radiation decreases with $a^{-4}$. Therefore we can write:
\be
\rho_\phi(T)\ = \ \rho_\phi(T_{in})\left(\frac{T}{T_{in}}\right)^3\sim
m_\phi^2\phi_{in}^2\left(\frac{T_0}{\sqrt{m_\phi M_P}}\right)^3
\ee
If the field $\phi$ is stable, these
oscillations will dominate the energy density of the universe and may
overclose it. Imposing that $\rho_\phi(T_0)< \rho_{critical}= 3H_0^2
M_P^2\sim (10^{-3} \rm{eV})^4$,
where $T_0, H_0$ are the temperature and Hubble parameter today,
puts a constraint on $\phi_{in}$,  $\phi_{in}<
10^{-10}\left(\frac{m_\phi}{100{\rm GeV}}\right)^{-1/4}M_P$. That is,
for $\phi_{in}\sim M_P$ a stable scalar field of mass $m_\phi > 10^{-26}$
eV will overclose the universe.

If the scalar field decays, which  is the most common situation,
another problem arises. Since the field couples with gravitational strength, its
decay will happen very late in the history of the universe and may
spoil nucleosynthesis.  This can be quantified as follows. The scalar field $\phi$
decays at a
temperature $T_D$ for which $H(T_D)\sim \Gamma_\phi$. Therefore using $\Gamma_\phi\sim m_\phi^3/M_P^2$ and the FRW
equations for $H\sim \Gamma_\phi$:
\be
\Gamma_\phi^2\sim \left(\frac{m_\phi^3}{M_P^2}\right)^2\sim
\frac{\rho_\phi(T_D)}{M_P^2}=\frac{\rho_\phi(T_{in})}{M_P^2}\left(\frac{T_D}{T_{in}}\right)^3
\ee
Using this and $\rho_\phi(T_{in})\sim m_\phi^2 \phi_{in}^2$,
$T_{in}^2\sim m_\phi M_P$ we find the decay temperature $T_D\sim
m_\phi^{11/6}M_P^{-1/6}\phi_{in}^{-2/3}$. At the temperature $T_D$
the energy density $\rho_\phi(T_D)$ gets converted into radiation of
temperature
\be
T_{RH}\simeq \left(\rho_\phi(T_D)\right)^{1/4}\sim
\left(M_P\Gamma_\phi\right)^{1/2}\sim
\left(\frac{m_\phi^3}{M_P}\right)^{1/2}.
\ee
If $T_{RH} \lesssim 10$ MeV the decay products of $\phi$ will spoil the successful
predictions of nucleosynthesis. This puts a bound on $m_\phi$ of
$m_\phi \gtrsim 100$ TeV.   The decay of $\phi$ causes an increase
in the entropy given by:
\be
\Delta \ =\ \left(\frac{T_{RH}}{T_D}\right)^3\, \sim
\frac{\phi_{in}^2}{m_\phi M_P}
\ee
which for $\phi_{in}\sim M_P$ gives a very large entropy increase
washing out any previously generated baryon asymmetry.
Therefore the standard cosmological moduli problem forbids gravity
coupled scalars in the range $m_\phi \lesssim 100$ TeV.
We will reconsider this problem in the next subsection for the large
volume string models.

\subsection{Other Problems}

\begin{itemize}

\item{}
{\it Gravitino overproduction}.
One proposal to avoid the cosmological moduli problem is through a heavy modulus
scenario, where $m_{\phi} \sim 1000 \hbox{TeV}$ with $m_{3/2} \sim 30 \hbox{TeV}$ and
$m_{soft} \sim 1 \hbox{TeV}$. However in this case the moduli are much heavier than the
gravitino and the $\phi \to 2 \psi_{3/2}$ decay channel is open. It has
recently been pointed out \cite{hepph0602061, hepph0602081} that
in this case the moduli decay to gravitinos is unsuppressed
and can occur with $\mc{O}(1)$ branching ratio.
This naturally leads to an overproduction of gravitinos at low energies, which
interfere with the successful nucleosynthesis predictions.
This problem appears on top of the more standard gravitino problem, in
which to avoid thermal gravitino overproduction  the
reheating temperature most be smaller than $10^{9}$ GeV.

\item{}
{\it Dark matter overproduction}
Even in heavy moduli scenarios where the moduli mass is $m_{\phi} > 100 \hbox{TeV}$,
the reheating temperature is still very low, $T_{reheat} \sim \mc{O}(10 \hbox{MeV})$.
As the moduli mass is much greater than that of the soft terms, the moduli will also decay to
TeV-scale supersymmetric particles with $\mc{O}(1)$ branching ratios. The reheat temperature is much lower
than that of the susy freeze-out temperature, which is typically $T_{freeze-out} \sim m_{LSP}/20 \gtrsim
\mc{O}(10) \hbox{GeV}$. The standard thermal relic abundance computation for susy dark matter does not apply and
a stable LSP is heavily overproduced.

\item{}
{\it Baryogenesis}
Moduli decays reheat the universe, generating large amounts of entropy and diluting any primordial baryon asymmetry.
At high temperatures, there exist mechanisms to generate a baryon asymmetry: for example, the electroweak sphaleron transitions that occur at
$T \sim 100 \hbox{GeV}$ violate baryon number.
However, the low reheat temperatures from moduli decay imply baryogenesis must occur at low temperatures, without the aid of the high energy
baryon number-violating processes.

\item{}
{\it Overshooting problem}. Usually the physical minimum of the
scalar potential is only a local minimum.
The initial conditions may typically
be that the  energy is much larger than the barrier
separating this minimum from the overall (zero coupling/infinite
volume) minimum. The field may then roll through the local minimum and pass
over the
barrier. This was emphasised in reference \cite{bs}. This is a problem of initial
conditions. Detailed studies of the time evolution of the
scalar field, following from equation (\ref{tevol}) have concluded
that this problem is less severe than originally thought
\cite{kaloper, brusteindealwis}.
It appears that Hubble damping together with the different redshift properties
of kinetic and potential energy
can be enough to avoid the field overshooting and running to infinity. This is a model
dependent problem that we will not address further.

\item{}{\it Inflationary destabilisation}
In practical models of moduli stabilisation, the barrier height separating the true minimum from the infinite
runaway is comparable to the depth of the AdS minimum, which is $\lesssim m_{3/2}^2 M_P^2$.
The barrier height is a measure of the maximum scale at which inflation can take place, as if the inflationary energy
scale is above the barrier height the potential is unstable to decompactification. During the inflationary epoch
this gives a relationship $H \lesssim m_{3/2}$ \cite{KalloshLinde}, which suggests that either the gravitino mass was very large during inflation
$m_{3/2} \gg 1 \hbox{TeV}$, or that inflation took place at a very low energy scale $H \ll 10^{16} \hbox{GeV}$. If the potential is such that
the gravitino mass is $\sim 1 \hbox{TeV}$ during inflation, typical inflationary energy scales will destabilise the potential.

\item{}
{\it Temperature destabilisation}. Finite temperature effects can modify
the scalar potential in such a way that the local physical minimum
is washed out at finite temperature due to the $T^4$ contribution to the
scalar potential from the coupling of the modulus to a thermal matter bath.
In this case the field naturally
rolls towards its decompactified zero coupling limit as in the
overshooting problem.
If moduli fields couple to the observable sector, the free energy of a
hot gas of observable particles contribute to the moduli potential
since moduli correspond to gauge couplings in the effective theory.
Since the free energy goes like $T^4$, for high enough temperatures this could destabilise the
zero-temperature minimum. The critical temperature was found to be of
order $10^{13}$ GeV \cite{buchmuller}.  If inflation occurs at energies
above $10^{15}$ GeV, there is no time for observable matter to be in
thermal equilibrium and the problem disappears \cite{graham}.
Then for small enough reheating temperature this is not a serious problem.

\end{itemize}

\section{Large Volume Moduli in the Early Universe}
\label{secLVMearly}

\subsection{Cosmological Moduli Problem}

Let us reanalyse the cosmological moduli problem for each of the
moduli fields present in the large volume models. In total there are
three classes of moduli: the complex structure and dilaton, the heavy
K\"ahler moduli and the light K\"ahler modulus. Let us discuss each
case on the basis of the analysis of the previous section.

\begin{enumerate}

\item{\it Complex structure and dilaton moduli}.
These fields have masses of order $20$ TeV and couple with
gravitational strength. In principle these are in the dangerous zone
for the CMP. However, as emphasised in \cite{hepth0505076}, the potential for
these fields dominates  the overall energy density, leading to runaway behaviour,
unless they sit at their
minimum. The reason is that for large volumes, their contribution to
the scalar potential is positive and suppressed only by $1/\mc{V}^2$, in
contrast to the K\"ahler moduli contribution that goes like
$1/\mc{V}^3$ at large volume.
Therefore such fields are naturally trapped at (or very close) to their
minimum early in the history of the universe and are not expected to have
dangerous oscillations ($\phi_{in} \lll M_P$).

\item{}
{\it Heavy moduli.}
The heavy moduli have masses of order $1000 $ TeV and
are coupled to matter at the string scale
($M_s \sim 10^{11} \hbox{GeV}$) rather than the Planck
scale $M_P \sim 10^{18} \hbox{GeV}$). They are therefore free from the
CMP as their
lifetime is extremely short, with $\tau \sim 10^{-17}$s.
Their decays will reheat the universe
to
\be T_{RH}\sim (M_P\Gamma_\Phi)^{1/2}\sim\left(M_P
m_{_\Phi}/M_s^2\right)^{1/2} m_{_\Phi}\sim 10^7\hbox{GeV}.
\ee
Furthermore, as the couplings of these moduli to the gravitini are Planck suppressed rather than
string suppressed, gravitino decay modes have tiny branching ratios. For example, the $\Phi \to 2 \psi_{3/2}$
decay mode occurs with a branching ratio of $\sim 10^{-30}$, in contrast to the $\mc{O}(1)$ expectations of
\cite{hepph0602061, hepph0602081}.

As the reheat temperature is high, it is
possible to start a Hot Big Bang at a relatively high
temperature, with the possibility of a conventional treatment of susy decoupling and axion evolution.
For the above reasons such moduli are very attractive for
reheating the universe after inflation.

\item{}
{\it Light modulus.} This field has a mass of order $1$ MeV with
gravitational strength interactions and it is thus dangerous for the CMP.
Notice that standard inflation
can never address the CMP because there is no reason for the scalar field
to be at its minimum just after inflation. To solve this problem we
need to have either a trapping mechanism to keep the fields in or close to
their minima or alternatively a period of late inflation. The best option for
this is thermal inflation \cite{hepph9510204} that we will discuss next.

\end{enumerate}

\subsection{Thermal Inflation}

Thermal inflation is not just another particular choice of scalar field
and potential energy to give rise to slow-roll inflation at high
energies. Thermal inflation is rather a general class of models that tend
to induce a  short period of low-temperature inflation in a
natural way. It is not an alternative to slow-roll inflation to solve
the big-bang problems and produce the density perturbations, but instead
complements it with a short period of low energy inflation that can
dilute some relic particles.

Thermal inflation was proposed in \cite{hepph9510204}. The observation is that
in supersymmetric models there are many flat directions (such as
the string moduli and others) that are lifted after supersymmetry
breaking. A field with such a flat direction, which we denote by
$\sigma$, can have a vacuum expectation value ({\it vev})
 much larger than its mass. If this is the case $\sigma$ is called a
 `flaton' field (not to be confused with the  inflaton).

The cosmological implications of a flaton field are quite interesting.
If the flaton field is in thermal equilibrium with matter, there is a finite
temperature contribution to its scalar potential:
\be
V\ =\ V_0 + (T^2-m_\sigma^2)\, \sigma^2 + \cdots
\ee
where we have  expanded around a local maximum of $\sigma$ taken to be
at
$\langle\sigma\rangle=0$. This is a false vacuum at temperatures
$T>T_c=m_\sigma$. At these temperatures $\sigma$ will be trapped at
the origin.
The zero temperature minimum is at $\langle
\sigma\rangle\equiv M_*\gg m_\sigma$. At a particular temperature
$T\simeq V_0^{1/4}>T_c$, the potential energy density $V_0$ starts to
dominate over the radiation energy $\sim T^4$ and a short period of
inflation
develops.
Inflation ends at $T=T_c$ when the field $\sigma$ becomes tachyonic at
the origin and runs towards its zero
temperature minimum. The number of efolds during this period of
inflation is $$N\sim \log\left(V_0^{1/4}/T_c\right)\sim \log\left(
M_*/m_\sigma\right)^{1/2},$$ where we have used that during inflation
the scale factor is inversely proportional to the temperature and
$V_0\simeq M_*^2
m_\sigma^2$. For $m_\sigma\sim 1$ TeV and $M_*\sim 10^{11}$ GeV, the
number of {\it e}-folds is $N\sim 10$. This is large enough to dilute the
surviving moduli and solve the cosmological moduli problem, but small
enough to not interfere substantially with the density perturbations
coming from the original period of inflation at higher energies.

It is interesting that the values preferred for the scales $M_*$ and
$m_\sigma$ are precisely the string and soft SUSY breaking scales in
our scenario.
It therefore
seems natural to try to implement thermal inflation in this scenario
with $M_*=M_s, m_\sigma\sim m_{3/2}$.
Candidate flaton fields can be any moduli with {\it vev} of order one
in string units and masses of the order of the soft masses. Singlet
open string modes abound in D-brane  constructions that have
precisely these properties. The heavy K\"ahler moduli also have the
right mass scale and {\it vev}. However their coupling to matter is suppressed
by the string scale and it is difficult for them be in thermal
equilibrium with observable matter.\footnote{This suppression is also present for the typical
  flaton fields considered in the literature. To be in thermal
  equilibrium with matter it is usually assumed that the flaton field
  couples to massive particles with a mass given by
  $\langle\sigma\rangle$. When  $\langle\sigma\rangle$ is close to
  zero these fields are light and allow $\sigma$ to be in thermal
  equilibrium.
 We may envisage a
  similar situation for the heavy moduli, as their vanishing implies a
  four-cycle collapsing and the appearance of extra massless fields. A proper
  treatment of this interesting
is beyond the low-energy effective action we have been
  using and would require further study.}
An explicit realisation of thermal inflation in our class of models is
beyond the scope of the present article, but it is encouraging to see
that they do have the right properties for thermal inflation to happen
with several candidate flaton fields. There is actually an explicit
candidate for thermal inflation using the properties of D-branes \cite{gia1}.

Other scenarios of low-temperature inflation could also work. Although
standard slow-roll inflation is difficult to obtain at low energies,
other variants such as locked inflation \cite{gia2} could be
promising, especially if they could be implemented within string theory.
A period of low-temperature inflation has also been proposed in
recent attempts to derive inflation from string theory \cite{bcsq}.

\subsection{Comparison with Other Scenarios}

Even though the moduli are generic in string compactifications their
physical implications change considerably depending on the details of
moduli stabilisation and supersymmetry breaking. At the moment there
are at least four main scenarios that can be distinguished:

\begin{enumerate}

\item{}
The generic gravity mediated scenario. In this case
all moduli are expected to get a mass proportional to the gravitino
mass. The argument is that their mass has to be proportional to the
auxiliary field that breaks supersymmetry divided by the strength of
the interaction that mediates the breaking of supersymmetry
($m\phi \sim \langle F\rangle/M_P$) which is precisely the gravitino
mass $m_{3/2}\sim 1$ TeV. All moduli are assumed to couple with
gravitational strength, and all moduli suffer from the cosmological moduli problem.

\item{}
Generic gauge mediated supersymmetry breaking. In this case $m_{3/2} \ll 1 \hbox{TeV}$. The moduli masses
are still of the same order of the gravitino mass, but now this may be as
low as $m_{3/2}\sim\langle F\rangle/M_P\sim 10^{-3}-10^3$ eV. They also
couple with gravitational strength and
induce a CMP even more severe than for the gravity-mediated scenario.

\item{} Mirage mediation \cite{hepth0411066}. This differs from conventional gravity mediation in
  the fact that the moduli masses are $m_\phi\sim m_{3/2}\,
  \log(M_P/m_{3/2})\sim 1000$ TeV. This improves on the CMP as moduli decay prior to BBN, but
  gives new problems with the overproduction of gravitini and susy
  dark matter as discussed above.

\item{}
Large volume models. In our case there are different classes of moduli.
The heavy moduli with $m_\Phi\sim 2 m_{3/2}\,
  \log(M_P/m_{3/2})\sim 1000$ TeV
 are free from both the CMP and gravitino
overproduction problems because their couplings are only suppressed by the string
scale. The light volume modulus has mass $\sim 1$ MeV
and couples gravitationally, and is subject to the CMP.
\end{enumerate}

The moduli spectrum for large-volume models does not remove all cosmological problems.
However, it does give quite different behaviour to more standard expectations.
One striking difference is the possibility of a high moduli reheating temperature, $T_{RH} \sim 10^7 \hbox{GeV}$, and the
commencement of a Hot Big Bang at a relatively early stage.
This arises because there exist moduli coupled to matter at the string, rather than the Planck, scale. In
the standard case where all moduli couple to matter at the Planck scale, the reheating temperature is invariably low.
For TeV scale moduli, $T_{RH} < 1 \hbox{MeV}$ and nucleosynthesis fails. Even in scenarios with heavy moduli,
with $m_{\Phi} \sim 1000 \hbox{TeV}$, the reheating temperature is still $T_{RH} < 1 \hbox{GeV}$.
High reheating temperatures are attractive because they can provide the necessary initial conditions for a period
of thermal inflation or for the standard susy relic abundance computation.

The other striking difference in the spectrum of the large-volume models is the volume modulus. This is extremely light ($\sim 1 \hbox{MeV}$)
and gravititationally coupled; such a field is unusual in models of gravity-mediated supersymmetry breaking.
Even if a Hot Big Bang has started at $10^7 \hbox{GeV}$, this field will subsequently
come to dominate the energy density of the universe
if its abundance is not diluted. This is why a period of
late-time (thermal) inflation may be necessary in order to dilute this volume modulus.
We now investigate the properties of this field in more detail.

\section{Large Volume Moduli in the Late Universe}
\label{secLateUniverse}

The combination of a light $\mc{O}(\hbox{MeV})$ modulus with gravity-mediated TeV-scale supersymmetry breaking is an
unusual and distinctive feature of the large-volume models, and offers the chance of obtaining a smoking-gun signal
for this class of models.
As the volume modulus
is stable on the lifetime of the universe, it may be present today as part of the dark matter.
As analysed in section \ref{sec3} above, it is unstable and may decay to $\gamma \gamma$ or, if kinematically accessible, $e^{+} e^{-}$.

We here analyse the possibilities for detecting these decays.
We first consider the photon flux due to $\chi \to \gamma \gamma$ decays, considering several
astrophysical sources. In section \ref{sec62} we generalise this to include
the dominant decay mode $\chi \to e^{+} e^{-}$,
and discuss
the relevance of this decay to the 511 keV positron annihilation line from the galactic centre.
 We start by leaving the lifetime, $\tau_{\chi}$, and mass, $m_{\chi}$ of the modulus unspecified:
these will subsequently be set as in section \ref{sec3} above.

\subsection{Photon flux from $\chi \to \gamma \gamma$ decays}

As sources, we consider the Milky Way halo, the diffuse background and nearby galaxy clusters.
We assume the field $\chi$ constitutes a fraction $\Omega_{\chi}/\Omega_{dm}$ of the dark matter.

\subsubsection*{The Milky Way Halo}

We assume the Milky Way halo to be spherical. For definiteness we consider two dark matter profiles,
isothermal and Navarro-Frenk-White (NFW), as these both allow an analytic treatment.
These are
\be
\rho_I(r) = \frac{\rho_0}{1 + \frac{r^2}{r_c^2}}, \qquad
\rho_{NFW}(r) = \frac{\rho_0}{\left(\frac{r}{r_s}\right) \left( 1 + \frac{r}{r_s} \right)^2}.
\ee
For both halo models, $\rho_0$, $r_c$ and $r_s$ are phenomenological parameters.
$r$ is measured from the galactic centre. By relating galactic coordinates $(x,b,l)$ to
Cartesian coordinates on the galactic centre, we can write
\bea
r^2 & = & (-R_0 + x \cos b \cos l)^2 + (x \cos b \sin l)^2 + (x \sin b)^2  \nonumber \\
\label{eqgc}
& = & (x - R_0 \cos b \cos l)^2 + R_0^2 (1 - \cos^2 b \cos^2 l).
\eea
Here $R_0 \sim 8 \hbox{kpc}$ is the distance of the sun from the galactic centre.
(\ref{eqgc}) allows the computation of $\rho(x, b, l)$ for any given halo model.

If the decay $\chi \to \gamma \gamma$ occurs at distance $x$ from a detector with cross-section
$\Delta_D$, the probability that a photon reaches the detector is
$
\mc{P} = \frac{\Delta_D}{4 \pi x^2} \ti 2,
$
where the factor of $2$ accounts for the two photons from the decay.
The number of photons arriving from distances
between $x$ and $x+dx$ in time $dt$ within a solid angle $d \Sigma$ is
\be
\underbrace{\frac{dt}{\tau_{\chi}}}_{\hbox{fractional decay probability}} \ti
\underbrace{n(\chi, x) \ti (x^2 dx) \ti d\Sigma}_{\hbox{no. of particles}} \ti
\underbrace{\frac{\Delta_D}{4 \pi x^2} \ti 2.}_{\hbox{arriving photons per decay}}
\ee
To obtain the total number of arriving photons, we integrate this quantity along  the radial ($x$) direction,
to obtain
\be
\mc{N}_{\gamma}(b, l) = \Delta_D \ti dt \ti \frac{2}{\tau_{\chi} m_{\chi}} \ti \frac{d \Sigma}{4 \pi}
\ti \left( \frac{\Omega_{\chi}}{\Omega_{dm}} \right) \int dx \, \rho(x).
\ee
We now perform the $\int dx \rho(x)$ integral for both the profiles considered.

\begin{enumerate}
\item Isothermal Profile

Here
\bea
\rho_I(x) & = & \frac{\rho_0 r_c^2}{r_c^2 + (x - R_0 \cos b \cos l)^2 + R_0^2(1 - \cos^2 b \cos^2 l)}.
\eea
Defining $R_{eff}^2 = r_c^2 + R_0^2(1 - \cos^2 b \cos^2 l)$,
we can do the integral using standard trigonometric substitutions, obtaining for the number of photons arriving per unit time
\be
N_{\gamma} = (\Delta_D) dt  \left( \frac{d \Sigma}{4 \pi} \right) \frac{2}{\tau_{\chi} m_{\chi}}
\left( \frac{\Omega_{\chi}}{\Omega_{dm}} \right) \rho_0 r_c^2 \left[ \frac{1}{R_{eff}} \left( \frac{\pi}{2} +
\arctan \left( \frac{R_0 \cos b \cos l}{R_{eff}} \right) \right) \right]
\ee
These photons are all mono-energetic of energy $\frac{m_{\chi}}{2}$ and will appear as a monochromatic line of
width $\Delta E$, the energy resolution of the detector at $E \sim \frac{m_{\chi}}{2}$. The intensity of this line is
\be
I_{line}(b, l) = \frac{N_{\gamma}(b, l)}{\Delta E}.
\ee

\item Navarro-Frenk-White Profile

For this case the integral $\int dx \, \rho(x)$ is performed in the appendix.
The resulting
number density of arriving photons is given by
\be
N_{\gamma}(b, l) = \Delta_D dt \left( \frac{d \Sigma}{4 \pi} \right) \frac{2}{\tau_{\chi} m_{\chi}} \left( \frac{\Omega_{\chi}}{\Omega_{dm}}
\right) \rho_0 r_s^3 X(b, l),
\ee
where
\bea
X(b, l) & \equiv & \frac{1}{r_s^2 - R_1^2(b,l)} \left( -1 - \frac{R_0^2 - R_1^2(b,l)}{R_0 + r_s} \right) \nonumber \\
& & -
\frac{r_s}{(r_s^2 - R_1^2(b,l))^{3/2}} \ln \left[ \frac{r_s R_0 + R_1^2(b,l) - \sqrt{(r_s^2 - R_1^2(b,l))(R_0^2 - R_1^2(b,l))}}
{R_1(b,l)(r_s + R_0)} \right] \nonumber \\
& & + \frac{r_s}{(r_s^2 - R_1^2(b,l))^{3/2}}
\ln \left[ \frac{R_1(b,l)}{r_s - \sqrt{r_s^2 - R_1^2(b,l)}} \right],
\eea
with $R_1(b,l) = \sqrt{R_0^2 (1 - \cos^2 b \cos^2 l)}$.
As before,
\be
I_{\gamma}(b, l) = \frac{N_{\gamma}}{\Delta E}.
\ee

\end{enumerate}
For numerical evaluations we use for the isothermal profile $\rho_0 = 7.8 \rm{GeV cm}^{-3}$ and $r_c = 2 \rm{kpc}$,
whereas for the NFW profile we use \cite{astroph0307026} $\rho_0 = 0.23 \rm{GeV cm}^{-3}$,
$r_s = 27 \rm{kpc}$, in both cases corresponding to $\rho(R_0) = 0.46 \rm{GeV cm^{-3}}$.

The galactic centre region, near $(b,l) = (0,0)$ is one of the most intensively observed areas of the galaxy and should contain
an excess of dark matter. It should therefore provide the best sensitivity in a search for a gamma-ray line due to modulus decay.
Integrating over a region $-15^\circ < b < 15^\circ, -15^\circ < l < 15^\circ$ for an NFW profile, we find a total photon flux of
\be
\label{jceq}
\mc{N}_{\gamma} = \left( \frac{\Omega_{\chi}}{\Omega_{dm}} \right) \left( \frac{6.5 \times 10^{25}s}{\tau_{\chi \to \gamma \gamma}} \right)
\left( \frac{2 \hbox{MeV}}{m_{\chi}} \right) \ti \left( 2.9 \ti 10^{-2} \hbox{photons cm}^{-2} \hbox{s}^{-1} \right).
\ee
The isothermal profile gives similar results. The INTEGRAL upper bound on $\sim 1 \hbox{MeV}$
gamma-ray lines from the galactic centre is that the line strength be
$\lesssim 5 \ti 10^{-5} \hbox{photons cm}^{-2} \hbox{s}^{-1}$ \cite{astroph0310793, astroph0604277},
so the absence of any such line constrains
\be
\frac{\Omega_{\chi}}{\Omega_{dm}} \lesssim 10^{-3} \left( \frac{2 \hbox{MeV}}{m_{\chi}} \right)^{2}.
\ee

\subsubsection*{Diffuse Background Emission}

Moduli decays across the history of the universe also contribute to the diffuse photon background.
We again relegate the computational details to the appendix, where we show
that the resulting photon flux intensity is
\be
\label{DecayDiff}
I_{\gamma}(E) = \frac{d \Sigma}{4 \pi} \ti \Delta_D \ti dt \ti d E_{\gamma}
\ti \left( \frac{\Omega_{\chi}}{\Omega_m} \right) \frac{2 \rho_0}{\tau_{\chi} m_{\chi}}
E_{\gamma}^{\half} \left( \frac{2}{m_{\chi}} \right)^{3/2} f\left( \frac{E'}{E_{\gamma}} \right) \frac{c}{H_0},
\ee
with $c$ the speed of light and
$$
f(x) = \left[ \Omega_m + \frac{1 - \Omega_m - \Omega_{\Lambda}}{x} + \frac{\Omega_{\Lambda}}{x^3} \right]^{-\half}.
$$
$\tau_{\chi}$ is the modulus lifetime and $\rho_0$ the current dark matter density. $E' \equiv \frac{m_{\chi}}{2}$ is
the original decay energy of the photons.

Because of the assumptions of homogeneity and isotropy, this
quantity will have the same value irrespective of direction. In figure \ref{DiffuseFlux}, we plot this quantity together with a fit to
the extragalactic diffuse gamma-ray background observed by COMPTEL. For $800 \hbox{keV} < E_{\gamma} < 30 \hbox{MeV}$ this is fit
by \cite{KappadathThesis}
\be
\label{ComptelDiff}
I_{\gamma}(E) = \left( \frac{E}{5 \hbox{MeV}} \right)^{-2.4} \ti \left( 1.05 \ti 10^{-4} \hbox{ photons cm}^{-2} \hbox{s}^{-1}
\hbox{sr}^{-1} \hbox{MeV}^{-1} \right).
\ee
\FIGURE{\makebox[15cm]{\epsfxsize=15cm \epsfysize=10cm
\epsfbox{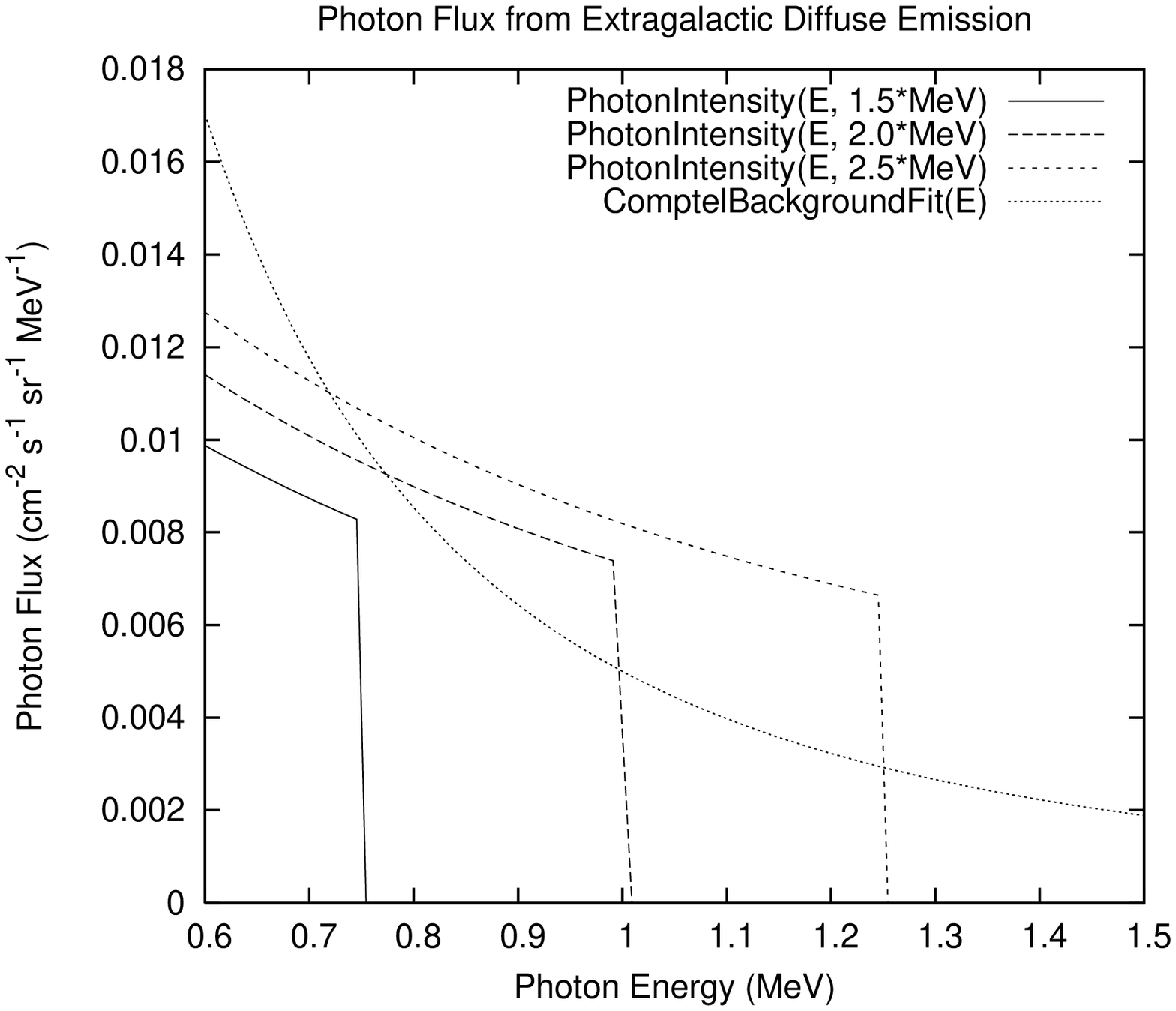}
}\caption{The extragalactic diffuse photon flux arising from moduli decays through the history of the universe.
We plot the flux arising for $\left( \frac{\Omega_{\chi}}{\Omega_{dm}} \right) = 1$ for moduli masses
$m_{\chi} = 1.5, 2$ and $2.5 \hbox{MeV}$.
We use the results of (\ref{cggcoup}) for the coupling of $\chi$ to photons.
As comparison we also plot a fit to the extragalactic diffuse gamma-ray background
observed by COMPTEL.}\label{DiffuseFlux}}
We see that for $m_{\chi} \gtrsim 1 \hbox{MeV}$ the combination of (\ref{DecayDiff}) and (\ref{ComptelDiff})
constrains the allowed $\chi$ density to be
\be
\frac{\Omega_{\chi}}{\Omega_{m}} \lesssim \left( \frac{1 \hbox{MeV}}{m_{\chi}} \right)^{3.5}.
\ee

\subsubsection*{Galaxy Clusters}

We can also consider specific local galaxy clusters.
A galaxy cluster is a locally overdense region of the sky, at a specific distance $D$ from the earth. We denote the
total dark mass of the cluster by $M$, with a fraction $\Omega_{\chi}/\Omega_{m}$ consisting of moduli.
The total number of moduli is then
$
\left( \frac{M}{m_{\chi}} \right) \left( \frac{\Omega_\chi}{\Omega_m} \right),
$
and thus the total number of arriving photons is
$$
\frac{\Delta_D}{4 \pi D^2} \ti 2 \ti \left( \frac{M}{m_{\chi} \tau_{\chi}} \right) \left( \frac{\Omega_\chi}{\Omega_m} \right).
$$
The photons give a monochromatic line of intensity
$$
I_{\gamma} = \frac{\Delta_D}{4 \pi D^2} \ti 2 \ti \left( \frac{M}{m_{\chi} \tau_{\chi}} \right)
\left( \frac{\Omega_\chi}{\Omega_m} \right).
$$
The line is redshifted from $E = \frac{m_{\chi}}{2}$ according to the distance of the cluster.
Considering for example the Coma or Perseus galaxy clusters, we find
\bea
I_{cluster} & \sim & 5 \ti 10^{-6} \, \hbox{photons} \, \rm{cm}^{-2} s^{-1} \left( \frac{1.5 \, \rm{MeV}}{m_{\chi}} \right)
\left( \frac{1.4 \ti 10^{26}s}{\tau_{\chi \to \gamma \gamma}} \right) \left( \frac{\Omega_{\chi}}{\Omega_m} \right) \\
& = & 5 \ti 10^{-6} \, \hbox{photons} \, \rm{cm}^{-2} s^{-1} \left( \frac{1.5 \, \rm{MeV}}{m_{\chi}} \right)^2
 \left( \frac{\Omega_{\chi}}{\Omega_m} \right).
\eea
This does not provide a competitive constraint on the modulus density.

\subsection{$\chi \to e^{+}e^{-}$ decays and the 511keV line}
\label{sec62}

If kinematically accessible, the light modulus $\chi$ can also decay to $e^{+}e^{-}$ pairs with a
decay rate
\be
\Gamma_{\chi \to e^{+}e^{-}} = \frac{m_e^2 m_{\chi}}{48 \pi M_P^2} \left( 1 - \frac{4 m_e^2}{m_{\chi}} \right)^{3/2}.
\ee
For masses $m_{\chi} \gtrsim 1 \hbox{MeV}$, this is the dominant decay mode. A plot of the relative branching fractions for
$\gamma \gamma$ and $e^{+}e^{-}$ are shown in figure \ref{decaymodes}. The decays of such a modulus could then be observed through the
positrons produced in the decay.
\FIGURE{\makebox[15cm]{\epsfxsize=15cm \epsfysize=10cm
\epsfbox{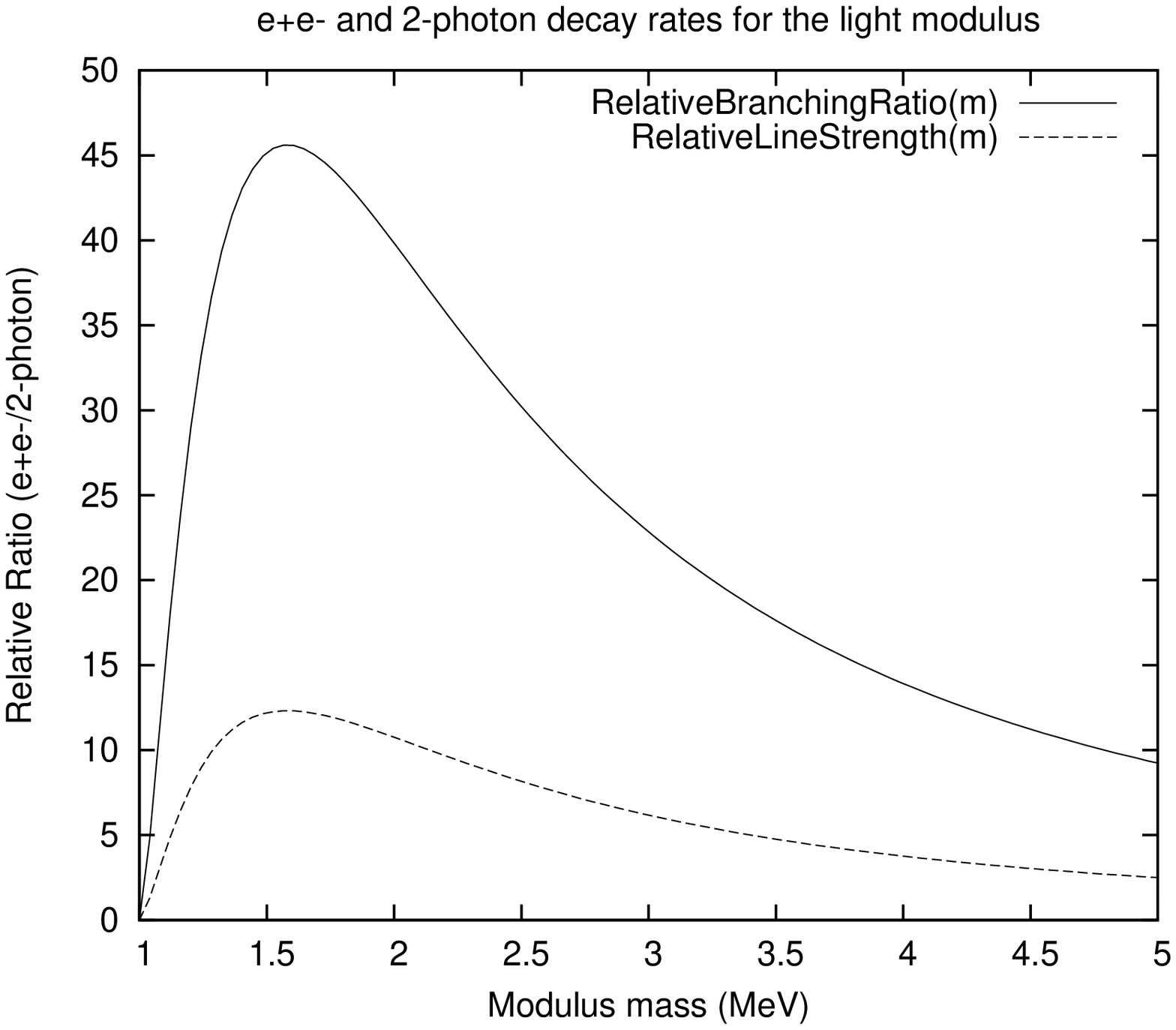}
}\caption{The relative branching ratio $Br(\chi \to e^{+}e^{-})/Br(\chi \to 2 \gamma)$
and line strengths arising from $\chi \to e^{+}e^{-}$ and $\chi \to 2 \gamma$
decay modes. The $e^{+}e^{-}$ decay mode dominates within the interesting range of moduli masses. For the $\chi \to
e^{+}e^{-}$ decay mode, `line strength' corresponds to that of the 511keV line from positron annihilation, taking into account
$3\gamma$ positronium decay. }\label{decaymodes}}

There exists an excess of 511keV photons from the galactic centre, which has been
detected for many years \cite{Johnson72, Leventhal78}. On the earth the flux of 511keV photons is
measured by the SPI spectrometer on the INTEGRAL satellite to be
$(0.96 \pm 0.06) \ti 10^{-3} \hbox{photons cm}^{-2} \hbox{ s}^{-1}$ \cite{astroph0702621}
This flux originates from the Milky Way bulge and it appears difficult for traditional astrophysical sources to account
for the intensity and location of the line.
The origin and injection energy of the positrons is unknown.
Positrons are produced relativistically and lose energy through ionisation and diffusion pocesses.
Eventually they become non-relativistic, annihilating with electrons to produce gamma rays.
The strongest constraint on the injection energy of the positrons
comes from
inflight annihilation of energetic positrons
to generate diffuse gamma rays above 511 keV.
The lack of an excess in the diffuse gamma ray spectrum
requires the positron injection energy to be $\lesssim 3 \hbox{MeV}$ \cite{astroph0512411} (more conservative models
may extend this to $\lesssim 10 \hbox{MeV}$ \cite{astroph0607374}).
The default assumption has to be that the positron excess arises from incompletely understood astrophysics associated with
e.g. the old stellar population of the galaxy. Nonetheless there remains the exciting possibility that the excess
has an exotic origin, such as arising from annihilating or decaying dark matter \cite{astroph0309686, hepph0402178, hepph0402220}.

In \cite{astroph0602296} decaying string moduli were considered as possible candidates to explain the 511keV line.
However, due to the above constraints on injection energies, $E_{inj} \lesssim 3(10) \hbox{MeV}$,
the modulus mass can only lie in a narrow window $1 \hbox{MeV} \lesssim m_{\chi}
\lesssim
6 \hbox{MeV} (20 \hbox{MeV})$.
Furthermore, the $2 \gamma$ decay must be considerably suppressed to avoid a photon line dramatically
exceeding that of the 511 keV line. In \cite{astroph0602296} it was necessary to impose both of the above
 constraints by hand.
In this respect the volume modulus $\chi$ that arises in the large-volume models is very appealing. The mass scale
arises in exactly the regime required, $\sim 1 \hbox{MeV}$, to avoid an overly high positron injection energy.
This mass scale is closely tied to the existence of TeV-scale supersymmetry and thus has a limited range
of allowed variation. The $2 \gamma$ decay rate also has a substantial parametric suppression in the coupling, by a
 factor $(\ln(M_P/m_{3/2}))^2$, compared to the $e^{+}e^{-}$ decay mode. Such a large suppression is essential: both
phase space effects and the $3 \gamma$ positronium decay make $e^{+}e^{-}$ decays less efficient at generating photons
than the direct $2 \gamma$ decay.

For the decays of the light modulus to saturate the 511keV line, we would require $\Omega_{\chi} \sim (\hbox{a few}) \ti 10^{-4}$,
depending on the details of the dark matter profile. This follows from an identical computation to (\ref{jceq}), except using the
$\chi \to e^{+}e^{-}$ decay rather than the $\chi \to \gamma \gamma$ decay model.\footnote{It is argued in \cite{astroph0507142} that for decaying dark matter to generate the 511keV line
a very cuspy dark matter profile would be necessary. However there are substantial astrophysical uncertainties
on the dark matter profile at small scales, and we also note that
as $\Omega_{\chi}/\Omega_{dm} \lesssim 10^{-3}$ the $\chi$ profile need not precisely coincide with the galactic dark matter profile.}
We can then use the computed couplings of the light modulus $\chi$ to electrons and photons to work out the relative intensity
of the 511 keV line from $\chi \to e^{+} e^{-} \to \ldots \to  e^{+} e^{-} \to \gamma \gamma$
and the monochromatic photon line arising from the direct decay of the modulus $\chi \to 2 \gamma$.\footnote{This contains the
implicit assumption that the locus of positron annihilations and the locus of modulus decays are approximately the same.}
The strength of the 511keV line depends on the fraction $f$ of $e^{+} e^{-}$ pairs that first form positronium,
which goes into $2\gamma$ and $3 \gamma$ final states with branching ratios $0.25:0.75$. Only the $2\gamma$ decay
generates 511keV line emission whereas the $3 \gamma$ decay gives continuum emission.
Assuming all positrons annihilate non-relativistically, the relative magnitudes of the
$3 \gamma$ positronium continuum and $2 \gamma$ direct annihilation fix the fraction of positrons that annihilate via positronium
to be $f = 0.97 \pm 0.02$ \cite{astroph0509298}. The number of photons produced per
$\chi \to e^{+}e^{-}$ decay is then $2((1-f)+0.25f) = 0.54$, in contrast to the 2 photons produced per $\chi \to \gamma \gamma$ decay.
The relative intensities of the 511keV line and the line from direct $2 \gamma$ decay is then
\be
\mc{R} = \frac{Br(\chi \to e^{+}e^{-})}{Br(\chi \to \gamma \gamma)} \ti \frac{0.54}{2}.
\ee
For the $\mbb{P}^4_{[1,1,1,6,9]}$ model analysed in detail in this paper,
we plot $\mc{R}$ as a function of $m_{\chi}$ in figure \ref{decaymodes}. We see that $\mc{R} \lesssim 12$, which would correspond to a line
intensity from direct decay of $\sim 8 \ti 10^{-5} \hbox{photons cm}^{-2} \hbox{s}^{-1}$. INTEGRAL constrains the strength of new gamma-ray
lines of $E_{\gamma} \sim 1 \hbox{MeV}$ from the galactic center to be $\lesssim 5 \ti 10^{-5} \hbox{photons cm}^{-2} \hbox{s}^{-1}$ \cite{astroph0310793, astroph0604277}, and so the existence of such a line is marginally ruled out.

There are however two caveats. The first is that the couplings have been computed in the high-energy theory, and so are valid
at the scale $m_s \sim 10^{11} \hbox{GeV}$. These should properly be renormalised down to the scale $m_{\chi} \sim 1 \hbox{MeV}$.
By analogy with the running of gauge couplings, or soft masses in supersymmetric scenarios, this may introduce $\mc{O}(1)$ corrections to
the high-scale coupling constants given in (\ref{cggcoup}) and (\ref{eecoupling}).
The precise results of the renormalisation depends on the details of the charged matter at $E \gtrsim 1 \hbox{TeV}$.
However, we can see that such corrections could
easily reduce the strength of the $\gamma \gamma$ line below the observable limit.

The second caveat is that while the coupling (\ref{eecoupling}) to electrons is model-independent - i.e. independent of the choice
of Calabi-Yau geometry and the details of the moduli stabilisation - the coupling (\ref{cggcoup})
 to photons is not and is specific to the $\mbb{P}^4_{[1,1,1,6,9]}$ model considered here.
This is because the derivation of the coupling to electrons depended only on the powers of volume present in the matter
kinetic terms. This volume scaling can be determined independently of any assumptions about the geometry of the
small cycle and precisely where the Standard Model is realised. In contrast, the coupling to photons depends on the extent to which the
light $\chi$ modulus has a component along the small cycle on which the Standard Model is supported. This is more dependent
on the details of the moduli stabilisation and on the geometry used to realise the Standard Model.

As the exclusion is only marginal it therefore remains possible that
the $e^{+}e^{-}$ decays of the volume modulus could saturate the 511keV flux while the $2 \gamma$ decays are below the
currently observable limit. Clearly any future observation of a new gamma ray
photon line from the galactic centre would greatly clarify this issue.

We also note that as the mass of the volume modulus is $\sim 1 \hbox{MeV}$ it is certainly possible that
$m_{\chi} < 1 \hbox{MeV}$. In this case the $e^{+}e^{-}$ decay is not kinematically accessible and there
is no possibility of accounting for the 511keV line through $\chi$ decays.

\section{Conclusions}

In this paper we have analysed the spectrum, couplings, decay modes and branching ratios of the moduli
for large-volume IIB string models. The supersymmetry breaking moduli divide into two classes with quite distinct properties.
\begin{enumerate}
\item
The first class consists of heavy moduli, with a mass $m_{\Phi} \sim 2 \ln(M_P / m_{3/2}) m_{3/2}$. For a
realistic supersymmetry breaking scale $m_{soft} \sim m_{3/2}/\ln(M_P/m_{3/2}) \sim 1 \hbox{TeV}$,
this mass is $\sim 1000 \hbox{TeV}$. These moduli are coupled to matter
at the string scale ($\sim 10^{11} \hbox{GeV}$) rather than the Planck scale ($\sim 10^{18} \hbox{GeV}$). They decay very rapidly,
with a lifetime $\tau \sim 10^{-17}$s, reheating to a temperature $\sim 10^7 \hbox{GeV}$. The couplings of such moduli to
to gravitini are
Planck-suppressed rather than string-suppressed.
The heavy modulus decay $\Phi \to \psi_{3/2} \psi_{3/2}$ has a branching ratio $\sim 10^{-30}$, and these moduli suffer neither
from the cosmological moduli problem nor the gravitino overproduction problem.
\item
The second class consists of the modulus controlling the overall volume. This has a mass $m \sim m_{3/2}
\left(\frac{m_{3/2}}{M_P}\right)^{\half}$. With TeV-supersymmetry breaking, this corresponds to
$m_{\chi} \sim 1 \hbox{MeV}$. This field couples to matter at the Planck scale rather than the string scale, and is stable
on the lifetime of the universe, with a decay lifetime $\tau \sim 10^{26}$s.
The existence of a light $\sim 1 \hbox{MeV}$ gravitationally coupled scalar is an extremely robust and model-independent
prediction of the large-volume scenario.
\end{enumerate}
The spectrum and properties of these moduli fields are quite distinct to those encountered
in other scenarios of supersymmetry breaking. In particular, the combination of a light $1 \hbox{MeV}$
modulus with gravity mediated supersymmetry breaking is a very distinctive prediction of the large-volume models.

While the analysis focused on one particular large-volume model, we
anticipate that this division will be more general, with
many small heavy moduli $\Phi_i$ and one light
volume modulus $\chi$.\footnote{For fibrations, there may also exist additional
light moduli with $m \sim m_{3/2} \left( \frac{m_{3/2}}{M_P} \right)^{2/3}$ \cite{Michele}.}
 The overall
volume dependence of the couplings are expected to be general whereas the
numerical coefficients will vary from model to model,
depending on the geometry of the corresponding Calabi-Yau manifold.
We also concentrated on the coupling of the moduli to QED since
electrons, positrons and photons are the only allowed decay channels
for the light modulus $\chi$ (neutrino interactions are
suppressed by $m_{\nu}$).
The coupling of $\chi$ to QCD and baryonic matter can still be interesting
to put constraints on the effects $\chi$ may have at low
energies. This can be done on the same lines as \cite{wise}, where a
careful analysis was made for all masses of dilaton-like
particles like $\chi$. The range of masses of $\chi$ is too low to
allow decays into baryonic matter and too large to be severely constrained by
violations of the equivalence principle.

We analysed the effect of the above moduli spectrum on early-universe cosmology. While this moduli spectrum does not solve
all cosmological problems, it does provide a quite different cosmological evolution. As the heavy moduli
are coupled to matter at the string scale, they decay very rapidly,
creating an early Hot Big Bang with a temperature $T \sim 10^7 \hbox{GeV}$. The reheating associated with
these moduli does not come with a gravitino
overproduction problem, as the gravitino is only Planck-scale coupled whereas ordinary matter is string-scale coupled.
However, the light modulus is subject to the cosmological moduli problem and
will tend to overclose the universe unless its abundance is diluted.
To dilute this abundance a period of thermal or alternative low-temperature inflation would seem necessary;
it may be possible to achieve this using the high temperatures generated by the decay of the heavy moduli.

We also analysed the astrophysical consequences of the light modulus, assuming it to be present today as part of the dark matter.
If this is the case this field may be detected through its decays to $2 \gamma$ or $e^{+}e^{-}$.
We computed its couplings and showed that its branching ratio to photons is parametrically
suppressed by a factor $\ln(M_P/m_{3/2})^2$ compared to the branching
ratio to $e^{+}e^{-}$. As the $e^{+}e^{-}$ branching ratio
is dominant, this opens the intriguing possibility that the decays of this field are responsible for the 511keV line
observed from the galactic centre.
In this regard the scale of the light modulus at $\mc{O}(1) \hbox{MeV}$ is attractive,
as the positron injection energies are required from astrophysics to be
$\lesssim 3 \hbox{MeV}$.
Using the tree-level high-scale couplings for the $\mbb{P}^4_{[1,1,1,6,9]}$ model, this possibility is marginally excluded, as we showed the
$2 \gamma$ decay line would then be slightly stronger than the observational bounds.
However, these couplings will be affected both by renormalisation to the low scale and by differing
 numerical coefficients that depend on the detailed geometry of the precise Calabi-Yau used, and so
the possibility that the decays of the light modulus is responsible
for the 511 keV line therefore remains open.

Finally, due to the $(\ln(M_P/m_{3/2}))^2$ magnitude of the
suppression for the monochromatic $2\gamma$ decay line,
it may be argued that if the 511 keV line is due to the $e^+e^-$ decay
of the light modulus, the monochromatic $2\gamma$ line should be within
reach in the near future.
It is also interesting to notice that the lifetime of the light
modulus is in the range that is close to being constrained by current CMB data
\cite{kamion} and future observations of the Hydrogen 21 cm line \cite{veint}.

\acknowledgments
We are grateful for conversations with S. Abdussalam, C. Boehm,
C.P. Burgess, M. Cicoli, J. Cline, G. Dvali, A. Fabian, R. Kallosh,
A. Linde, M. Pospelov, S. Sarkar, K. Suruliz.
JC is funded by Trinity College, Cambridge. FQ is partially funded by
PPARC and a Royal Society Wolfson award.

\begin{appendix}

\section{Moduli Kinetic Terms and Mass Matrices}

The kinetic terms for the moduli fields can be computed from the K\"ahler potential. As we work at large
volume, in computing the kinetic terms we drop the $\alpha'$ corrections and other terms that are suppressed
at large volume. Keeping the contributions to the metric at leading order in $\tau_b$, we have
\be
\mc{K} = -2 \ln \left( \frac{1}{9\sqrt{2}} \left( \tau_b^{3/2} - \tau_s^{3/2} \right) + \frac{\xi}{2 g_s^{3/2}} \right),
\ee
giving
\be
\mc{K}_{i \bar{j}} = \left( \begin{array}{ccc} \mc{K}_{b \bar{b}} & &\mc{K}_{b \bar{s}} \\
\mc{K}_{s \bar{b}} & &\mc{K}_{b \bar{s}} \end{array} \right) =
\left( \begin{array}{ccc} \frac{3}{4 \tau_b^2} & & -\frac{9 \tau_s^{\half}}{8 \tau_b^{5/2}} \\
-\frac{9 \tau_s^{\half}}{8 \tau_b^{5/2}} & & \frac{3}{8 \tau_s^{\half} \tau_b^{3/2}} \end{array} \right),
\ee
and
\be
\mc{K}^{-1}_{i \bar{j}} = \left( \begin{array}{ccc} \frac{4
    \tau_b^2}{3} & & 4 \tau_b \tau_s \\
4 \tau_b \tau_s & & \frac{8 \tau_b^{3/2} \tau_s^{1/2}}{3} \end{array} \right).
\ee
In deriving this it is necessary to recall that $\frac{\partial}{\partial \tau_b} = \frac{1}{2} \frac{\partial}{\partial T_b}$.
To compute the mass matrix we need to evaluate the second derivatives of the potential at the minimum.
We start with
$$
V = \frac{a_s^2 \lambda \sqrt{\tau_s} e^{-2 a_s \tau_s}}{\tau_b^{3/2}} - \frac{\mu a_4 \tau_s e^{-a_4 \tau_s} \vert W_0 \vert}{\tau_b^3}
+ \frac{\nu \vert W_0 \vert^2}{\tau_b^{9/2}}.
$$
It can be shown after some computation that at the minimum of this potential,
\bea
\label{app4}
e^{-a_s \tau_s} & = & \left( \frac{\mu}{2 \lambda} \right) \frac{\vert W_0 \vert}{\tau_b^{3/2}}
\frac{\sqrt{\tau_s}}{a_s} \left( 1 - \frac{3}{4 a_s \tau_s} -
\frac{3}{(4a_s\tau_s)^2}+ \ldots \right), \nonumber \\
\tau_s^{3/2} \left( \frac{\mu^2}{4 \lambda} \right) & = & \nu \left( 1
+ \frac{1}{2 a_s \tau_s} + \frac{9}{(4a_s\tau_s)^2}+ \ldots \right).
\eea
Using the results of (\ref{app4}), we can show that to second order in an expansion
in $\epsilon =1/(4a\tau_s)$:
\bea
\frac{\partial^2 V}{\partial \tau_b^2} & = & \frac{9 \vert W_0 \vert^2 \nu}{2 \tau_b^{13/2}} \left( 1
+ \frac{1}{2a_s \tau_s} \right), \\
\frac{\partial^2 V}{\partial \tau_s^2} & = & \frac{2 a_s^2 \vert W_0
  \vert^2 \nu}{\tau_b^{9/2}} \left( 1 - \frac{3}{4 a_s \tau_s} + \frac{6}{(4a_s\tau_s)^2}\right), \\
\frac{\partial^2 V}{\partial \tau_s \tau_b} & = & -\frac{3 a_s \vert
  W_0 \vert^2 \nu}{\tau_b^{11/2}} \left( 1 - \frac{5}{4 a_s \tau_s} + \frac{4}{(4a_s\tau_s)^2}\right).
\eea
The mass matrix is given by $M_{i \bar{j}} = \frac{1}{2} V_{i \bar{j}}$. This is
\be
M_{i \bar{j}} = \left( \begin{array}{cc} \frac{9 \vert W_0 \vert^2 \nu}{4 \tau_b^{13/2}} \left( 1
+ \frac{1}{2a_s \tau_s} \right) & -\frac{3 a_s \vert
  W_0 \vert^2 \nu}{2 \tau_b^{11/2}} \left( 1 - \frac{5}{4 a_s \tau_s} + \frac{4}{(4a_s\tau_s)^2}\right) \\
  -\frac{3 a_s \vert
  W_0 \vert^2 \nu}{2 \tau_b^{11/2}} \left( 1 - \frac{5}{4 a_s \tau_s} + \frac{4}{(4a_s\tau_s)^2}\right) &
  \frac{a_s^2 \vert W_0
  \vert^2 \nu}{\tau_b^{9/2}} \left( 1 - \frac{3}{4 a_s \tau_s} + \frac{6}{(4a_s\tau_s)^2}\right) \end{array} \right),
\ee
and so
\be
\label{matrizA}
\mc{K}^{-1} M^2 = \frac{2 a_s\langle \tau_s \rangle |W_0|^2 \nu}{3 \langle \tau_b \rangle^{9/2} } \left( \begin{array}{ccc}
  -9(1-7\epsilon) & & 6 a_s \langle \tau_b \rangle(1-5\epsilon+16\epsilon^2)\\
-\frac{6\langle \tau_b \rangle^{1/2}}{\langle \tau_s \rangle^{1/2}} (1-5\epsilon+4\epsilon^2) & &  \, \,\,\,
\frac{4 a_s \langle \tau_b \rangle^{3/2}}{\langle \tau_s \rangle^{1/2}}  (1-3\epsilon+6\epsilon^2) \end{array} \right).
\ee

\section{Integrals}

\subsection{NFW Halo}

For the NFW halo, the integral $\int dx \rho(x)$ we wish to perform is
\be
\label{nfwint}
\rho_0 r_s^3 \int_0^{\infty} dx \frac{1}{\sqrt{(x-\alpha)^2 + \beta^2} (r_s + \sqrt{(x-\alpha)^2 + \beta^2})^2},
\ee
where
\bea
\alpha & = & R_0 \cos b \cos l, \nonumber \\
\beta^2 & = & R_0^2 (1 - \cos^2 b \cos^2 l ). \nonumber
\eea
We can rewrite this as
$$
\int_{-\alpha}^{\infty} dy \frac{1}{\sqrt{y^2 + \beta^2} (r_s + \sqrt{\beta^2 + y^2})^2}.
$$
We split this integral up:
$$
= \int_{-\alpha}^0 dy \frac{1}{\sqrt{y^2 + \beta^2} (r_s + \sqrt{y^2 + \beta^2})^2} +
\int_{0}^{\infty} dy \frac{1}{\sqrt{y^2 + \beta^2} (r_s + \sqrt{y^2 + \beta^2})^2}.
$$
We now let $y^2 + \beta^2 = z^2$, so
$$
y = \Big\{ \begin{array}{c} - \sqrt{z^2 - \beta^2} \qquad y<0 \\ \sqrt{z^2 - \beta^2} \qquad y>0 \end{array}.
$$
The integral then becomes
$$
= \int_\beta^{\sqrt{\alpha^2 + \beta^2}} \frac{dz}{\sqrt{z^2 - \beta^2} (r_s+z)^2} + \int_\beta^{\infty}
\frac{dz}{\sqrt{z^2 - \beta^2} (r_s+z)^2}.
$$
Writing $z' = z+r_s$, we obtain
$$
= \int_{\beta+r_s}^{r_s+\sqrt{\alpha^2 + \beta^2}} \frac{dz'}{z'^2 \sqrt{z'^2 - 2 r_s z' + (r_s^2 - \beta^2)}} +
\int_{\beta+r_s}^{\infty} \frac{dz'}{z'^2 \sqrt{z'^2 - 2 r_s z' + (r_s^2 - \beta^2)}}.
$$

This is now a standard integral which can be found in integral tables. The indefinite integral ($\alpha'$ and $\beta'$ have no relation
to $\alpha$ and $\beta$ above)
$$
\int \frac{dz'}{z'^2 \sqrt{z'^2 + \alpha' z' + \beta'}} =
- \frac{\sqrt{z'^2 + \alpha' z' + \beta'}}{\beta' z'}
+ \frac{\alpha'}{2 \beta' \sqrt{\beta'}} \ln \left[ \frac{2 \sqrt{\beta' (z'^2 + \alpha' z' + \beta')} + \alpha' z' + 2 \beta'
}{z'} \right].
$$
After some manipulation, the integral (\ref{nfwint}) becomes $\rho_0 r_s^3 X(b, l)$, where
\bea
X(b, l) & \equiv & \frac{1}{r_s^2 - R_1^2(b,l)} \left( -1 - \frac{R_0^2 - R_1^2(b,l)}{R_0 + r_s} \right) \nonumber \\
& & -
\frac{r_s}{(r_s^2 - R_1^2(b,l))^{3/2}} \ln \left[ \frac{r_s R_0 + R_1^2(b,l) - \sqrt{(r_s^2 - R_1^2(b,l))(R_0^2 - R_1^2(b,l))}}
{R_1(b,l)(r_s + R_0)} \right] \nonumber \\
& & + \frac{r_s}{(r_s^2 - R_1^2(b,l))^{3/2}}
\ln \left[ \frac{R_1(b,l)}{r_s - \sqrt{r_s^2 - R_1^2(b,l)}} \right],
\eea
and $R_1(b,l) = \sqrt{R_0^2 (1 - \cos^2 b \cos^2 l)}$.

\subsection{Diffuse Background Emission}

Diffuse emission arises from moduli decays throughout the history of the universe.
At a time $t$ before the present, the scale factor of the universe satisfied
$\frac{a_t}{a_0} = \frac{1}{1+z}$.
The moduli number density was thus larger,
$
n_t(\chi) = n_0(\chi)(1 + z)^3,
$
and the number of decay events between time $t$ and time $t+ dt$ was
$$
N = n_t(\chi) \frac{dt}{\tau_{\chi}} = n_0(\chi)(1 + z)^3 \frac{dt}{\tau_{\chi}}.
$$
The photons produced by these decay redshift and now have energies between $E_{\gamma}$
and $E_{\gamma} + d E_{\gamma}$, with
$
E_{\gamma} = \frac{m_{\chi}}{2(1+z)} \equiv \frac{E'}{1+z},
$
where we define $E' = \frac{m_{\chi}}{2}$.
As the universe is expanding, there is a current number density of photons
$$
N_{\gamma}(E_{\gamma}) dE_{\gamma} = 2  n_0(\chi) (1+z)^3 \frac{dt}{\tau_{\chi}} \ti \frac{1}{(1+z)^3}
$$
of photons with energies between $E_{\gamma}$ and $E_{\gamma} + d E_{\gamma}$.
Now, as
$
E_{\gamma} = \frac{E'}{1+z},
$
we have
$$
d E_{\gamma} = - \frac{E'}{(1+z)^2} \frac{dz}{dt} dt.
$$
The relation between redshift $z$ and time $t$ is determined by the matter and dark
energy content of the universe. For matter density $\Omega_m$ and
dark energy $\Omega_{\Lambda}$, we have
\bea
\frac{dt}{dz} & = & - \frac{1}{H_0} \frac{1}{(1+z)^{5/2}} \left[ \Omega_m + \frac{1 - \Omega_m - \Omega_{\Lambda}}{1+z}
+ \frac{\Omega_{\Lambda}}{(1+z)^3} \right]^{-\half} \\
& = & - \frac{1}{H_0} \frac{1}{(1+z)^{5/2}} f(1+z),
\eea
where
$
f(x) = \left[ \Omega_m + \frac{(1 - \Omega_m - \Omega_{\Lambda})}{x} + \frac{\Omega_{\Lambda}}{x^3}\right]^{-\half}.
$
We can therefore write
$
dz = - \frac{H_0 dt (1+z)^{5/2}}{f(1+z)}.
$
Using $\frac{E'}{E_{\gamma}} = 1+z$, we thus have
\be
dt = \frac{E_{\gamma}^{\half} f \left(\frac{E'}{E_{\gamma}}\right)}{(E')^{3/2} H_0} dE_{\gamma}.
\ee

This relates the range in departure times of the photons and the energy differences they have now.
Then
$$
N_{\gamma}(E_{\gamma}) dE_{\gamma} = \frac{2 n_0(\chi)}{\tau_{\chi}} \frac{E_{\gamma}^{\half} f(E'/E_{\gamma}) d E_{\gamma}}{H_0 (E')^{3/2}}
\hbox{ photons with energies between $E_{\gamma}$ and $E_{\gamma} + d E_{\gamma}$.}
$$
The energy density per unit volume in decay photons is
\be
\rho_{\gamma}(E_{\gamma}) dE_{\gamma}= \frac{2 n_0(\chi)}{\tau_{\chi}} \left( \frac{2 E_{\gamma}}{m_{\chi}} \right)^{3/2}
f \left( \frac{E'}{E_{\gamma}} \right) \frac{dE_{\gamma}}{H_0},
\ee
with a photon number density
\be
N_{\gamma}(E_{\gamma}) = \frac{\rho_{\gamma}(E_{\gamma})}{E_{\gamma}}.
\ee
This converts into a photon flux at a detector observing a solid angle $d \Sigma$ of
\be
\mc{N} = (d \Sigma) \ti \Delta_D \ti (c dt) \ti \frac{N_{\gamma}}{4 \pi}.
\ee
If we observe a solid angle $d \Sigma$, the number of photons arriving in time $dt$ from the diffuse cosmic
background is then
\be
\mc{N} = \frac{1}{4 \pi} (d \Sigma) \ti (\Delta_D) \ti (dt) \ti \frac{2 n_0(\chi)}{\tau_{\chi}}
E_{\gamma}^{\half} \left( \frac{2}{m_{\chi}} \right)^{3/2} f \left( \frac{E'}{E_{\gamma}} \right)
\frac{c}{H_0} d E_{\gamma}.
\ee
This now gives us the number of arriving photons, $\rm{s}^{-1} \rm{sr}^{-1} \rm{cm}^{-2} (\rm{keV})^{-1}$.
We can if we wish write this in terms of $\Omega_{\chi}$, using
$$
n_0(\chi) = \left( \frac{\Omega_{\chi}}{\Omega_m} \right) \frac{\rho_0}{m_{\chi}},
$$
to obtain
$$
I_{\gamma}(E) = \frac{d \Sigma}{4 \pi} \ti \Delta_D \ti dt \ti d E_{\gamma}
\ti \left( \frac{\Omega_{\chi}}{\Omega_m} \right) \frac{2 \rho_0}{\tau_{\chi} m_{\chi}}
E_{\gamma}^{\half} \left( \frac{2}{m_{\chi}} \right)^{3/2} f\left( \frac{E'}{E_{\gamma}} \right) \frac{c}{H_0}.
$$
$\rho_0$ is the present day dark matter density.
\end{appendix}

\end{document}